\documentclass[accepted=2023-06-17,a4paper]{quantumarticle}
\pdfoutput=1 
\usepackage[T1]{fontenc}
\usepackage{array,cite}
\usepackage{amsmath}
\usepackage{amssymb}
\usepackage{graphicx,hyperref}
\usepackage{wasysym}
\usepackage[utf8]{inputenc}
\usepackage[english]{babel}
\usepackage[T1]{fontenc}
\usepackage{amsmath}
\usepackage[numbers,sort&compress]{natbib}

\makeatletter

\pdfpageheight\paperheight
\pdfpagewidth\paperwidth

\providecommand{\tabularnewline}{\\}

\usepackage{physics,bbm}
\usepackage{braket}

\def\uno{\mathbbm{1}}

\def\sZ{\mathcal{Z}}

\newcommand{\bk}[2]{\braket{#1|#2}} 

\usepackage[export]{adjustbox}% http://ctan.org/pkg/adjustbox
\let\oldincludegraphics\includegraphics
\renewcommand\includegraphics[2][]{%
  \oldincludegraphics[#1,max width=\linewidth,max height=\textheight]{#2}
}

\makeatother

\usepackage{babel}
\begin{document}
\title{Fast quantum transfer mediated by topological domain walls}
\author{Juan Zurita}
\email{juzurita@ucm.es}
\affiliation{Instituto de Ciencia de Materiales de Madrid (CSIC), Cantoblanco,
E-28049 Madrid, Spain}
\affiliation{Departamento de F\'isica de Materiales, Universidad
Complutense de Madrid, E-28040 Madrid, Spain}

\author{Charles E. Creffield}
\affiliation{Departamento de F\'isica de Materiales, Universidad
Complutense de Madrid, E-28040 Madrid, Spain}
\author{Gloria Platero}
\affiliation{Instituto de Ciencia de Materiales de Madrid (CSIC), Cantoblanco,
E-28049 Madrid, Spain}
\begin{abstract}
The duration of bidirectional transfer protocols in 1D topological models usually scales exponentially with distance. In this work, we propose  transfer protocols in multidomain SSH chains and Creutz ladders that lose the exponential dependence, greatly speeding up the process with respect to their single-domain counterparts, reducing the accumulation of errors and drastically increasing their performance, even in the presence of symmetry-breaking disorder.
We also investigate how to harness the localization properties of the Creutz ladder---with two localized modes per domain wall---to choose the two states along the ladder that will be swapped during the transfer protocol, without disturbing the states located in the intermediate walls between them.
This provides a 1D network with all-to-all connectivity that can be helpful for quantum information purposes.

% Transfer protocols in 1D topological models are usually implemented
% between the left and right end modes of the system, and their speed
% often exponentially reduces with transfer distance. In this work,
% we propose a way to harness the localization properties of a multidomain
% Creutz ladder, a flat-band topological model with two protected states
% per domain wall, to choose the two modes along the ladder which will
% be swapped using a transfer protocol. Only protected states are involved
% in the transfer, and so it is robust against symmetry-preserving disorder.
% Remarkably, one protected state per domain wall is left undisturbed,
% even if it is located between the two swapped states. An effective
% 1D chain of protected states is then established, where any pair of
% them can be swapped without affecting the others, providing a versatile
% and resilient platform for quantum information purposes. Additionally,
% we propose transfer protocols in multidomain Creutz ladders and SSH
% chains which exponentially speed up the process with respect to their
% single-domain counterparts, thus reducing the accumulation of errors
% and drastically increasing their performance, even in the presence
% of symmetry-breaking disorder.
\end{abstract}
\maketitle

\section{Introduction}
We are currently experiencing the noisy, intermediate-scale era of
quantum computing (NISQ) \cite{Preskill2018}, where the first signs of
quantum advantage---the ability of quantum computers to outperform classical ones for certain tasks---start to appear. The path ahead towards large error-correcting
codes and, eventually, fault-tolerant quantum computation will most
likely require a wide array of tools and techniques to fight decoherence.
One of these techniques can be the use of symmetry-protected states
in topological insulators, which are especially robust against some
types of noise, and are already available in the laboratory.

These topological boundary states naturally decouple from the bulk
of the system, and thus form an invariant subspace of the Hamiltonian
around zero energy, which has support on the boundaries of the material.
This allows us to define quantum information protocols in 2D \cite{Yao2013,Dlaska2017,Lemonde2019,Bello2017a}
and 1D \cite{Leijnse2011,Almeida2016,Lang2017,Jin2017,Estarellas2017,Mei2018,Boross2019,Yuce2019,Dangelis2020,Tan2020,Comaron2020,Hu2020,Cao2021,Chen2021,Palaiodimopoulos2021,Han2021,Yuan2021}
which transport a particle from one region of the boundary to another,
with little to no loss to the bulk, even in noisy regimes. These \emph{particle}
transfer protocols can then be used in different ways to implement
\emph{state} transfer protocols \cite{Lang2017,Mei2018,Boross2019}, that is, to transport encoded quantum information.
%, or building a quantum computer with specific couplings so that its qubit states are equivalent to the empty ($\ket{0}$) and filled ($\ket{1}$) states of a topological Hamiltonian \cite{Mei2018}\textendash{} or other remote quantum operations \cite{Lang2017,Boross2019}.
In this paper we are mainly concerned with the particle transfer protocols
in the topological systems we study, leaving their specific applications
in quantum information protocols for future works.

We tackle a well-known issue in protocols based on the Rabi oscillations between exponentially decaying end modes, which is the \textendash also exponential\textendash{} increase in the time of transfer as a function of system length. These protocols are bidirectional, in contrast with other options like topological pumping, and are induced by the hybridization of the end modes. We show that a solution for this scaling problem can be found in multidomain models, in which the protected states in domain walls along the system can act as signal amplifiers, exponentially decreasing the transfer time between distant states even with the most simple control protocols. This solution boils down to the fact that a process akin to $N$ sequential transfers between domain walls is much faster than a single transfer between the two ends, i.e., that $N e^{L/N} \ll e^L$.

Firstly, we propose exponentially accelerated protocols between the ends of multidomain SSH chains, where each domain wall can hold
one boundary state. Domain walls in the SSH chain and related models
have been studied since its inception \cite{Su1980,Munoz2018,Longhi2019a,Dangelis2020,Qi2021},
and other transfer protocols using them exist in the literature \cite{Longhi2019a},
but our proposal and scaling results have not been yet explored, as
far as we know. Our protocols thus provide the first bidirectional
topologically protected transfers whose duration does not scale exponentially with distance.

We also investigate the possibilities of particle control
provided by magnetic interference in the Creutz ladder (CL) \cite{Creutz1999},
a quasi-1D topological insulator which was recently realized in three
different cold atom systems \cite{Song2018,HyounKang2020,He2021},
and can also be implemented with state-of-the-art technology in superconducting
circuits \cite{Alaeian2019a,Hung2021} (in which a CL plaquette has
been implemented) and photonic lattices \cite{Mukherjee2018}. In
a particular regime, an orthogonal basis of spatially compact energy
eigenstates can be found, something associated with the complete flattening
of both bands in momentum space. This phenomenon is called Aharonov-Bohm
caging \cite{Vidal1998,Creffield2010,Bello2016,Junemann2017}, and
it is caused by the destructive interference of paths due to the magnetic
Peierls phases. Flat band models are interesting playgrounds to study
the effects of interactions and disorder \cite{Tovmasyan2013a,Takayoshi2013a,Derzhko2015,Kuno2020c,Pelegri2020a,Roy2020,Cadez2021,Nakai2022,Kim2022},
and, like topological models, also show interesting properties for
quantum information purposes \cite{Rontgen2019,Kuno2020b}. In a
Creutz ladder with multiple topological domains, where each domain
wall can hold up to two protected states, we show that it is possible to decouple some of the boundary states of the system from all the others at will.

Remarkably, if this is done to a state in a two-state domain wall,
its partner can be used to leapfrog over it, thus allowing the transfer
of a particle through the wall while leaving the decoupled state undisturbed
and protected inside it. This allows for more complex transfer protocols between the ends and walls of the model. Even though the CL topological phase diagram
is well-known \cite{Li2015,Zurita2019}, the properties of its possible
domain walls have not been studied in the literature, to the best
of our knowledge.

In Section \ref{sec:SDoms}, we present the multidomain SSH chain, its protected states and bidirectional transfer protocols. Then, in Section \ref{sec:SFast}, we explore the exponential acceleration that can be achieved using domain walls as signal amplifiers, and the performance against disorder. Finally, in Section \ref{sec:Creutz} we study the states in Creutz ladder domain walls and explore their more complex transfer protocols.

\section{Multidomain SSH chain\label{sec:SDoms}}
\subsection{Topological domains and walls}

\begin{figure}[!tph]
\mbox{%
\includegraphics[width=1\columnwidth]{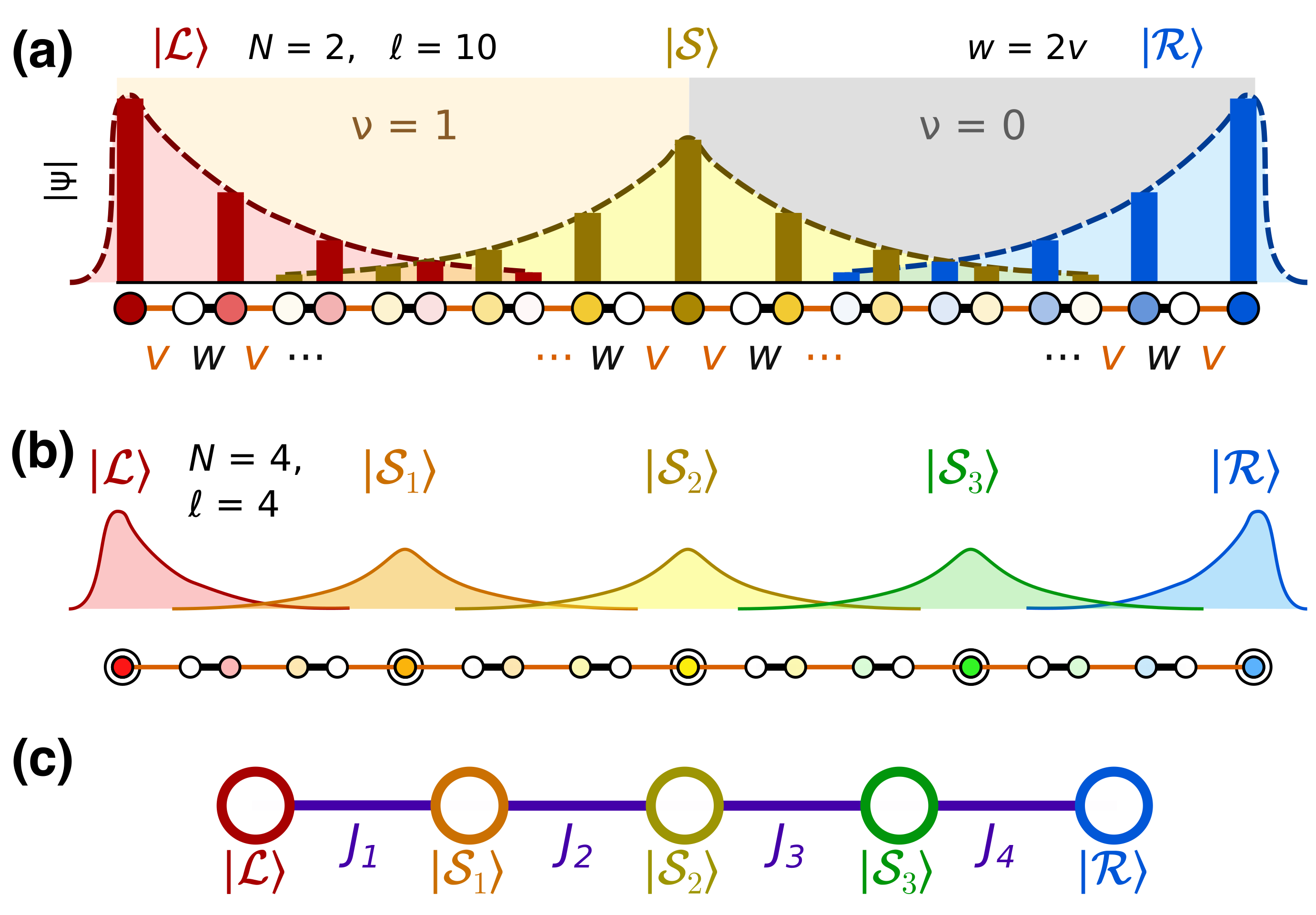}%
}\caption{(a) Two-domain SSH chain, with $\ell=10$ interior sites per domain ($L=23$). The wavefunctions of its three protected boundary states are depicted for $w=2v=1$. The right state $|\mathcal{R}\rangle$ does not follow the usual bulk-boundary correspondence because the last unit cell is not complete, but it is also protected. (b) Four-domain SSH chain with $\ell=4$ ($L=21$), and its five boundary states. (c) Effective model for the boundary states in the chain depicted in (b).\label{fig:001}}
\end{figure}

In this section, we consider a multidomain SSH chain. The SSH chain has intracell ($v$) and intercell ($w$) hopping amplitudes, and it will be trivial (topological) if $v>w$ ($v<w$), with a winding number of $\nu=0$ ($1$). Different domains can be established by setting up contiguous spatial regions with different dimerizations, as shown in Fig. \ref{fig:001} (a,b).

The Hamiltonian that describes an SSH chain with $N$ such domains
is:
\begin{align}
\mathcal{H} & \!=\!-\!\!\sum_{k=1}^{N}\!\left\{\! \!\delta_{1,k\!\!\!\!\!\mod\!2}\!\!\sum_{x=x_{k}^{i}}^{x_{k}^{f}}\!\!\left(v_{k}c_{x,b}^{\dagger}c_{x,a}\!+\!wc_{x+1,a}^{\dagger}\!c_{x,b}+\!h.c.\right)\right.\nonumber \\
 & \left.+\delta_{0,k\!\!\!\!\!\mod\!2}\!\!\sum_{x=x_{k}^{i}}^{x_{k}^{f}}\!\!\left(v_{k}c_{x-1,b}^{\dagger}c_{x,a}\!+\!wc_{x,a}^{\dagger}c_{x,b}\!+\!h.c.\right)\right\} , \label{eq:H_SSH}
\end{align}
where $x_{k}^{i}=\left\lceil ((k-1)(\ell+1)/2\right\rceil +1$ (resp.
$x_{k}^{f}=\left\lceil k(\ell+1)/2\right\rceil $) is the first (resp.
last) unit cell that the $k$-th domain has support on, with $\lceil\cdot\rceil$ being the ceiling function, $\ell$ is
the number of inner sites in a domain, $w$ takes a fixed value and
$v_{k}$ are the control parameters for each domain, and $\delta_{x,y}$
are Kronecker deltas. The first term is nonzero for odd domains, and
the second one is nonzero for even ones. Operator $c_{x,\alpha}^{(\dagger)}$
destroys (creates) a particle in unit cell $x=1,\ldots,\left\lceil L/2\right\rceil $
and sublattice $\alpha=a,b$. We restrict our study to the single-particle case. $L=N(\ell+1)+1$ is the total length
of the chain. We consider even values of $\ell$,
given that then, all boundary states are localized in a single site
in the fully dimerized limit. We also consider $w=1$ throughout.

Each of the $N-1$ domain walls holds an exponentially localized state $\ket{\mathcal{S}_{k}}$, which, in addition to the left and right states, make a total of $N+1$ protected boundary states. Each of them is chiral, with support only in one of the two sublattices ($a$ or $b$), as shown in Fig. \ref{fig:001} (a) for a three-domain chain. The bulk-boundary correspondence explains the left and domain wall states, the latter due to the fact that $|\nu_k-\nu_{k+1}|=1$. However, as seen in the figure, the rightmost domain of the chain can give rise to a right state even if the domain is trivial. This is due to its chiral basis: the end of a trivial system can be made topological by removing a site or vice versa. This is a well-known even-odd effect that has been studied since the conception of the SSH model \cite{Su1980,Ganeshan2013,Estarellas2017,Mei2018,Yuan2021}.

The boundary states in the multidomain SSH chain have the form:
\begin{align}
\ket{\mathcal{L}} & =-\mathcal{M_{L}}\sum_{x=1}^{1+\ell/2}\left(-\frac{w}{v}\right)^{-x}\ket{x,a}\label{eq:Lm-2-1}\\
\ket{\mathcal{R}} & =-\mathcal{M_{R}}\sum_{x=L-\ell/2}^{L}\left(-\frac{w}{v}\right)^{x-\left\lceil L/2\right\rceil -1}\ket{x,\Delta_{N}}\label{Rm-2-1}\\
\ket{\mathcal{S}_{k}} & =\mathcal{M}_{\mathcal{S}_{k}}\!\!\left[\ket{x_{0}^{(k)},\Delta_{k}}+\!\!\!\!\!\sum_{x=x_{0}^{(k)}-\ell/2}^{x_{0}^{(k)}-1} \!\! \left(-\frac{w}{v}\right)^{x-x_{0}^{(k)}}\ket{x,\Delta_{k}}\right.\nonumber \\
 & \left.+\sum_{x=x_{0}^{(k)}+1}^{x_{0}^{(k)}+\ell/2}\left(-\frac{w}{v}\right)^{x_{0}^{(k)}-x}\ket{x,\Delta_{k}}\right],\label{eq:Pm-2-1}
\end{align}

where $x_{0}^{(k)}=\left\lceil [k(\ell+1)+1]/2\right\rceil $ is the
unit cell where the $k$-th domain wall is, and $\text{\ensuremath{\Delta_{k}=a\,(b)}}$ if
$k$ is even (odd).

%#######################################################################################
%#######################################################################################

Each state is confined to its two adjacent domains (or one, in the case of end states). The states deviate slightly from the expressions above for very short domain lengths if $v/w\lesssim 1$, due to finite size effects ($\delta_{\mathcal{L}}=\norm{\ket{\mathcal{L}}_{\mathrm{num}}-\ket{\mathcal{L}}_{\mathrm{analyt}}}=0.18$
for $\ell=2$ and $v=3w/4$), but are fairly accurate otherwise ($\delta_{\mathcal{L}}=0.06$ for $\ell=4$ and $v=w/2$).

In the fully dimerized limit ($v=0$), all boundary states are completely localized in the isolated sites at each end or wall, decoupled from the rest of the system. We will use this limit at the beginning and end of our protocols, as is done in other works \cite{Lang2017}, in order to make the initial state easy to prepare, and the final state easy to work with (or read).

 %#######################################################################################
% \begin{figure}[!tph]
% \mbox{%
% \includegraphics[width=1\columnwidth]{NEWFIG.png}%
% }\caption{(a) Two-domain SSH chain, with $\ell=10$ interior sites per domain ($L=23$). The wavefunction of its three protected boundary states are depicted for $w=2v=1$. The right state $|\mathcal{R}\rangle$ does not follow the usual bulk-boundary correspondence because the last unit cell is not complete. (b) Four-domain SSH chain with $\ell=4$ ($L=21$), and its five boundary states. (c) Exact diagonalization results for the occupation number as a function of site index in a left-to-right transfer protocol in a two-domain chain with $\ell=4$ ($L=11$). Its control pulse is shown below. We use the site index $j$ instead of $x,\alpha$.  (d) Left-to-right transfer protocol in the four-domain ladder depicted in (b). The control pulses are now different for the outer and central domains.\label{fig:new}}
% \end{figure}
% %#######################################################################################

\subsection{Effective Hamiltonian}

We can obtain the Hamiltonian block that describes the evolution of the subspace spanned by the topological boundary modes, which is decoupled from the bulk states, in the basis formed by them:

\begin{equation}
\mathcal{H}_{\mathrm{eff}}\!=\!J_{1}\ketbra{\mathcal{S}_{1}}{\mathcal{L}}\!+\!J_{N}\ketbra{\mathcal{R}}{\mathcal{\mathcal{\mathcal{S}}}_{N-1}}\!+\!\!\sum_{k=2}^{N-1}\!\!J_{k}\ketbra{\mathcal{\mathcal{\mathcal{S}}}_{k}}{\mathcal{\mathcal{\mathcal{S}}}_{k-1}}\!+\!h.c.,\label{eq:HeffS}
\end{equation}

where $J_{1}=\bra{\mathcal{\mathcal{S}}_{1}}\mathcal{H}\ket{\mathcal{L}}$,
$J_{k}=\bra{\mathcal{S}_{k}}\mathcal{H}\ket{\mathcal{S}_{k-1}}$ for
$k=2,\ldots,N-1$, and $J_{N}=\bra{\mathcal{R}}\mathcal{H}\ket{\mathcal{S}_{N-1}}$. 

We will refer to these effective hopping amplitudes as "overlaps" throughout the paper for convenience, although the actual overlap of the states (e.g. $\bk{\mathcal{L}}{\mathcal{S}_1}$) is always zero due to their different chiralities. The fact that the $J_k$ are nonzero cause the boundary modes to hybridize, which in turn enables the transfer.

Using the analytical form for the states above, we obtain:
\begin{align}
J_{1} & =v\mathcal{M_{L}}\mathcal{M}_{\mathcal{S}_{1}}\left(-\frac{w}{v}\right)^{-\ell/2-1} \label{eq:t1}\\
J_{k} & =-v\mathcal{M}_{\mathcal{S}_{k}}\mathcal{M}_{\mathcal{S}_{k-1}}\left(-\frac{w}{v}\right)^{-\ell/2},\,\,\,\,\,k=2,\ldots,N-1\\
J_{N} & =v\mathcal{M_{R}}\mathcal{M}_{\mathcal{S}_{N-1}}\left(-\frac{w}{v}\right)^{-\ell/2-1}, \label{eq:tN}
\end{align}
where the normalization constants\footnote{In the case where the domains to the left ($l$) and right ($r$) of the wall have different $v$, the second formula takes the form $\mathcal{M}_{\mathcal{S}_{k}}=([1-(v_l/w)^2]^{-1}+[1-(v_r/w)^2]^{-1}-1)^{-1/2}$.} can be closely approximated for $\ell\gtrsim 3$ as:

\begin{align}
\mathcal{M}_{\mathcal{L}} & =\mathcal{M_{R}}=\sqrt{w^{2}/v^{2}-1}\\
\mathcal{M}_{\mathcal{S}_{k}} & =\sqrt{\frac{w^{2}-v^{2}}{w^{2}+v^{2}}}\,\,\,\,\,\,\,\forall k.
\end{align}

The effective Hamiltonian (\ref{eq:HeffS}) was also derived in \cite{Munoz2018}. We will refer to it as an effective model, although the only approximation made is the validity of expressions (\ref{eq:Lm-2-1}-\ref{eq:Pm-2-1}).

It takes the form of a 1D chain of length $N+1$, with only nearest-neighbour coupling. In Fig. \ref{fig:001} (c) we show the effective model corresponding to the four-domain chain in (b). Thus, the problem of implementing a transfer from $|\mathcal{L}\rangle$ to $|\mathcal{R}\rangle$ now takes the form of a transfer from the first site to the last in a chain of $N+1$ sites.

Regarding the validity of this effective model, it will be accurate as long as the bulk-boundary decoupling is preserved, that is, in levels of noise and disorder smaller than half the gap (as we discuss below), and for smaller time scales than the relaxation time of the system. The speed of the protocols is an advantage in this sense, and our protocols are, in principle, orders of magnitude faster than typically cited relaxation times in systems like weakly-interacting cold atoms \cite{Flesch2008} and semiconducting quantum dots \cite{Stano2022}.

%Its details are included in Appendix \ref{sec:SSH_Eff}

% %#######################################################################################
% \begin{figure}[!tph]
% \mbox{%
% \includegraphics[width=1\columnwidth]{NewEightScale.png}%
% }\caption{(a) Effective model for the protected manifold in the four-domain SSH chain depicted in Fig. \ref{fig:new} (b). (b) Occupation of the  
% topological states during an LR transfer in a two-domain SSH chain
% with domain length $\ell=4$ ($L=11$), $v_{\textrm{tr}}=0.5$ and $t_\textrm{prep}=15$., where the only $\mathcal{S}$
% state is represented in yellow. Numerical results for the full SSH
% chain Hamiltonian are pictured in continuous line, while the analytical
% prediction by the effective Hamiltonian is plotted in dotted line.
% (c) Occupation of the topological states in an LR transfer in the
% four-domain SSH chain pictured in Fig. \ref{fig:new} (b), with $\ell=4$ ($L=21$), $v_{\textrm{tr}}=0.5$ and $t_\textrm{prep}=15$. \label{fig:EffS}}
% \end{figure}
% %#######################################################################################

\subsection{Controlled transfer protocol}

As mentioned above, the initial (final) state of the transfer will have the particle localized in the leftmost (rightmost) site of the chain. This is done for convenience, and to stop the process at the correct time. The topological states can be adiabatically transformed into these states by slowly switching $v$ off. In order for the adiabatic condition to be satisfied, this preparation time must be larger than the characteristic time scale of the system $\tau=2/\Delta$, with $\Delta$ being the energy gap. The time scale $\tau$ is the inverse of the difference between the protected manifold and the bulk states\footnote{The reasons to use this time scale of the system and not the bonding energy of the end modes are discussed in \cite{Lang2017} and in Appendix \ref{sec:TimeEvo}}. For an SSH chain with $w=2v=1$, the point with the smaller gap in the protocol, this time scale is $\tau \sim 8$. This was checked numerically, obtaining results compatible with adiabatic preparation for preparation times of around $t_{\mathrm{prep}}\lesssim 10/w$. We include further details about the adiabatic time evolution in Appendix \ref{sec:TimeEvo}.

In a single-domain SSH chain, a transfer can be achieved by varying the value of $v$ in the following way:

\begin{equation}
v_{\mathrm{tr}}(t)\!=\!\begin{cases}
v_{\mathrm{tr}}\sin^{2}(\Omega t) & \textrm{for }0\leq t<t_{\mathrm{prep}}\\
v_{\mathrm{tr}} & \textrm{for }t_{\mathrm{prep}} \leq t<t_{\mathrm{tr}}\!-\! t_{\mathrm{prep}}\\
v_{\mathrm{tr}}\sin^{2}[\Omega(t-t_{\mathrm{tr}})] & \textrm{for }t_{\mathrm{tr}}-t_{\mathrm{prep}} \leq t\leq t_{\mathrm{tr}},
\end{cases}\label{eq:vCParam-1}
\end{equation}
where $v_{\mathrm{tr}}$ is the pulse amplitude, $t_{\textrm{tr}}$ is the total transfer time, and $t_{\mathrm{prep}}=\pi/(2\Omega)$. All times are expressed in units of $\hbar/J$, and we take $\hbar=1$. We use this type of pulse, which has been employed in the literature \cite{Longhi2019}, because it provides smooth
initial and final preparation stages, and also a period where
the control parameter is constant, thus simplifying its experimental implementation. More complex protocols acting on the topological states which may yield shorter transfer times can also be considered, and will be explored more thoroughly in future works.

When several domains are considered, it is necessary to change the height of the pulse in each domain $D$, $v_\textrm{tr}^{(D)}$, independently, in order to obtain the optimal transfer time.

The value of $v_\textrm{tr}^{(D)}$, in turn, controls the effective hopping amplitude $J_{k=D}$ of the effective chain, according to Eqs. (\ref{eq:t1}-\ref{eq:tN}). It also depends on the value of $v_\textrm{tr}^{(D\pm 1)}$, through the normalization constant of the $\mathcal{S}$ states.

By changing these parameters, it is possible to freely engineer each pulse. To find the optimal control parameter values for each $N$, we fixed $v_1=v_N=w/2$, and then searched numerically for the optimal values of $v_2=v_{N-1}$, $v_3=v_{N-2}$, etc., always keeping the system symmetric under spatial inversion to ensure symmetric effects on the left and right states, something necessary for bidirectional transfers. Typical ratios for $J_1/J_2$ in the midpoint of the protocols are around $0.83$ for the studied cases. The fidelity threshold required for the final state in all transfer simulations of the paper is taken to be $f_{0}=0.995$ unless stated otherwise, well above the current estimates of quantum error correction thresholds, which are around $f=0.990$ \cite{Knill2005,Wang2011,Fowler2012,Egan2020}.

We simulated left-to-right (LR) transfers numerically using exact diagonalization to solve the Schr{\"o}dinger equation for the tight-binding Hamiltonian in Eq. (\ref{eq:H_SSH}). We apply the time evolution operator $e^{-i\mathcal{H}(t)\Delta t}$ to the wavefunction $\ket{\psi(t)}$ to make the system evolve from time $t$ to $t+\Delta t$. We consider a time step of $\Delta t=0.1$ for all simulations, which has been numerically confirmed to provide the same result as shorter time steps even in the fastest protocols studied.

\begin{figure}[!tph]
\mbox{%
\includegraphics[width=1\columnwidth]{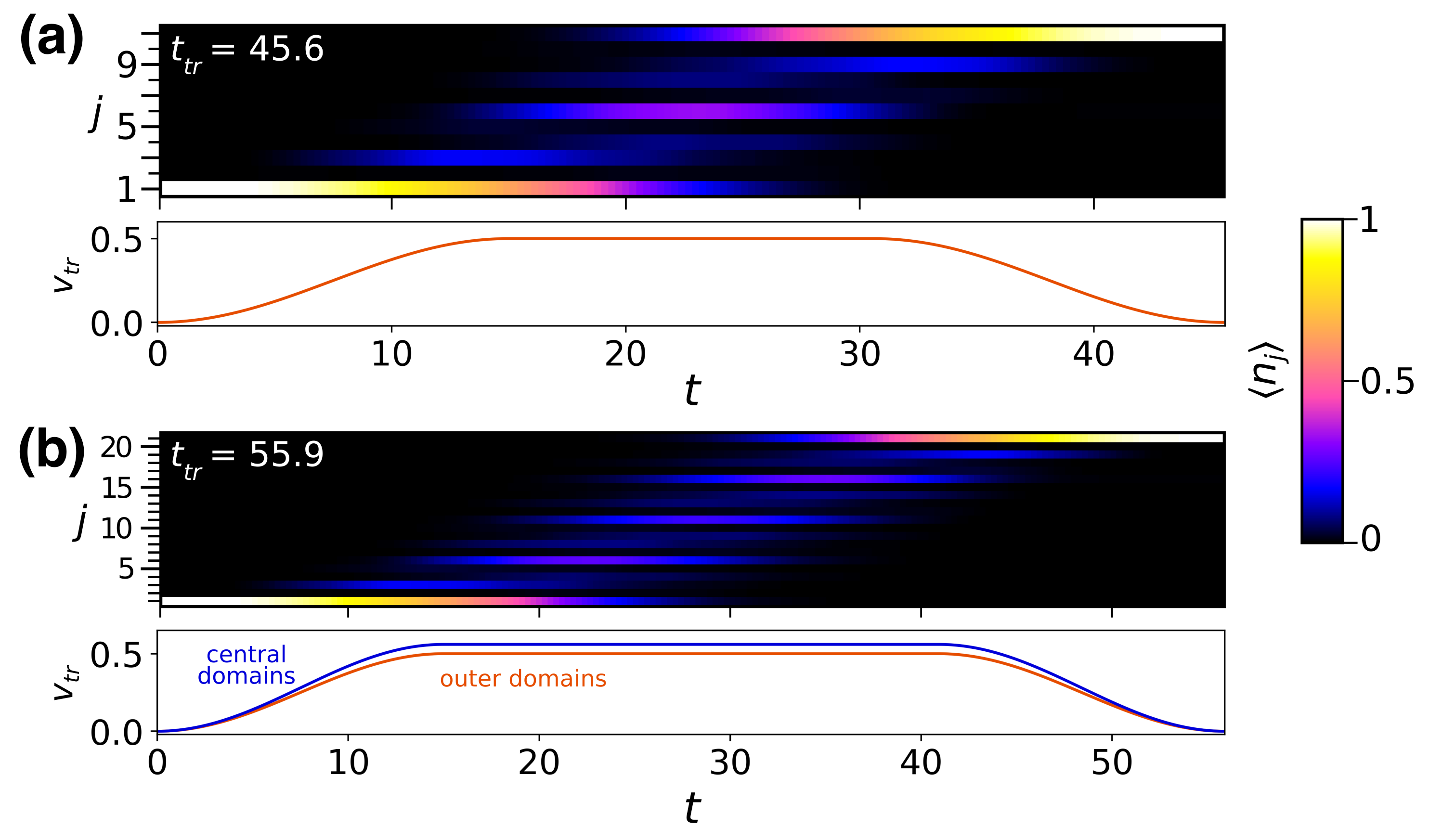}%
}\caption{(a) Exact diagonalization results for the occupation number as a function of site index in a left-to-right transfer protocol in a two-domain chain with $\ell=4$ ($L=11$). Its control pulse is shown below. We use the site index $j$ instead of $x,\alpha$.  (b) Left-to-right transfer protocol in the four-domain ladder with $\ell=4$ depicted in Fig. \ref{fig:001} (b). The control pulses are now different for the outer and central domains. All times are expressed in units of $\hbar/J$.\label{fig:002}}
\end{figure}

In Fig. \ref{fig:002} we show the results for LR transfers in chains with domains of length $\ell=4$, for a two-domain ($L=11$) and a four-domain ($L=21$) chain. We use $v_{\textrm{tr}}=w/2=0.5$ in the first one, and $v_{\textrm{tr}}^{(1)}=0.5$, $v_{\textrm{tr}}^{(2)}=0.56$ in the second one. The preparation time is $t_{\textrm{prep}}=15$ for both, and the transfer times are $t_{\textrm{tr}}=45.6$ and $55.9$, respectively. We consider $\ell=4$ because the decay length of the end modes for $w=2v$ is $\lambda = [\log (w/v)]^{-1} = 3.32$, and so their overlap is not too small. We show the time evolution during the transfer of the occupation at each site $\langle n_j\rangle $, using the site index $j(x,\alpha)=2x-\delta_{\alpha,a}$, where $x$ is the unit cell and $\alpha$ is the internal coordinate. 

%#######################################################################################
%#######################################################################################

\begin{figure}[!tph]
\mbox{%
\includegraphics[width=1\columnwidth]{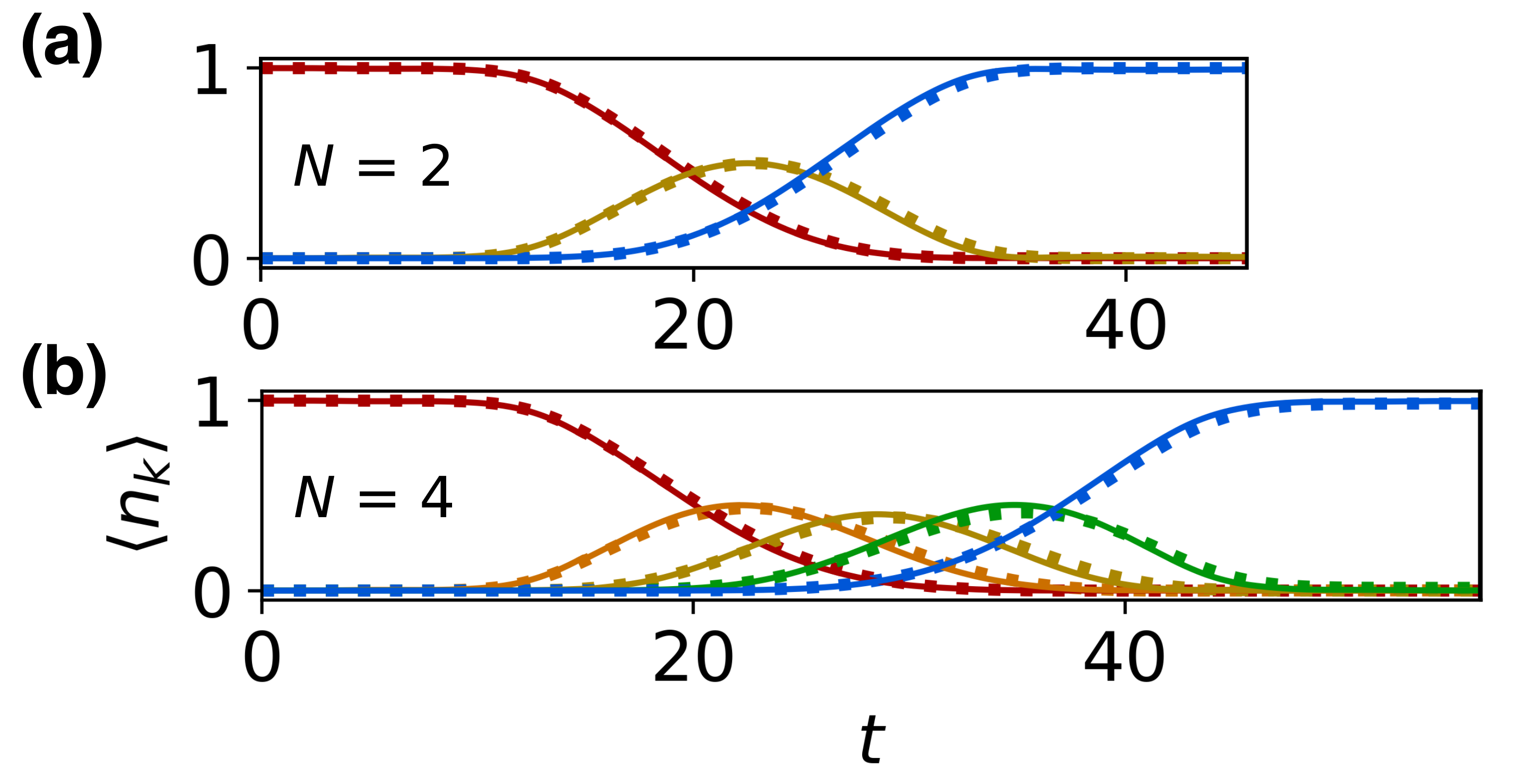}%
}\caption{(a) Occupation of the topological states during an LR transfer in a two-domain SSH chain with domain length $\ell=4$ ($L=11$), $v_{\textrm{tr}}=0.5$ and $t_\textrm{prep}=15$, where states $\mathcal{L},\mathcal{S}$ and $\mathcal{R}$ are represented in red, yellow and blue, resp. Numerical results for the full SSH
chain Hamiltonian are shown in continuous line, while the analytical
prediction by the effective Hamiltonian is plotted in dotted line.
(b) Occupation of the topological states in an LR transfer in the
four-domain SSH chain depicted in Fig. \ref{fig:001} (b), using the same color code, with $\ell=4$ ($L=21$), $v^{(1)}_{\textrm{tr}}=0.5$, $v^{(2)}_{\textrm{tr}}=0.56$ and $t_\textrm{prep}=15$. \label{fig:003}}
\end{figure}

In Fig. \ref{fig:003}, we plot the occupation of each of the boundary states obtained by exact diagonalization (solid line) during the same transfer protocols as in Fig. \ref{fig:002}, and compare it to the effective model prediction (dotted line). The same time scale has been used for both subplots, showing their similar duration. By choosing the appropriate pulses, an exponential speed-up can be obtained in cases with large $L$ when compared to the single-domain case, as we explore in Section \ref{sec:SFast}. Furthermore, as demonstrated in \cite{Apollaro2012,Banchi2013}, a high fidelity transfer protocol can be obtained for any length of the effective model, $N$, by modifying only $J_1,J_2,J_{N-1}$ and $J_N$, assuming perfect precision. Hence, in principle, our scheme could be extended for any number of domains.

A complex phase is acquired in general by the transferred components of the wavefunction. This phase factor can be obtained using the effective model, and has the form:

\begin{equation}
\zeta(\ell,n_{w})=\begin{cases}
(-1)^{(n_{w}+\ell)/2}i & \textrm{for even }n_{w}\\
(-1)^{(n_{w}+1)/2} & \textrm{for odd }n_{w},
\end{cases}
\end{equation}

where $n_w$ is the number of domain walls between the transferred
states and $\ell$ is the domain length.

Other transfer protocols have been proposed in multidomain SSH chains \cite{Boross2019,Almeida2016,Estarellas2017,Dangelis2020,Yuan2021}, but, to the best of our knowledge, they all consider only two consecutive domains. The protocols considered are also unidirectional, except in \cite{Estarellas2017}. While preparing this manuscript, we came across another work \cite{Almeida2016} that proposes transfer protocols using SSH-like segments joined by additional links and suggests a possible time advantage, although the topological properties of their model and protocols are not studied. The multidomain SSH chain has also been studied recently in \cite{Munoz2018}, although not in the context of transfer protocols.

%#######################################################################################
%#######################################################################################

\section{Fast long-range transfer\label{sec:SFast}}
\subsection{Topological domain walls as amplifiers}

Using the effective model we can derive a result that has been a known problem in topological bidirectional transfer protocols since their conception\footnote{To the best of our knowledge, all 1D bidirectional transfer protocols are driven by the time evolution of several hybridized modes. Their main advantage, compared to other possibilities like adiabatic passage \cite{Longhi2019a} or the movement of protected states \cite{Mei2018}, is that their bidirectionality
allows them to be used to implement a wider array of quantum operations
in an efficient manner, as explored in \cite{Lang2017} for remote quantum gates. Another
issue in some of the unidirectional protocols is the narrowing or
closing of the gap, which can be problematic for large distances,
and also restricts the transfer time.}. If we consider the transfer times for a single-domain SSH chain as a function of length, we find:

\begin{equation}
t_{\textrm{tr}}^{(N=1)}\!=\!\frac{\pi v}{2(w^{2}-v^{2})}\! \left(\frac{w}{v}\right)^{\ell/2+2}\!\!=\!\frac{\pi v}{2(w^{2}-v^{2})}\! \left(\frac{w}{v}\right)^{L/2+1}\label{eq:T_tr_SSH1}
\end{equation}

We can see that the duration of the protocol scales exponentially with the transfer distance $L$, if we keep the control parameter $v$ fixed. This is because the exponential localization of the modes causes
their overlap \textendash and bonding energy\textendash{} to shrink
exponentially as the distance between them increases. This causes
the transfer dynamics to be extremely slow in large systems, which is a general phenomenon in bidirectional topological protocols.

One possible solution to this problem is studied in \cite{Lang2017}: given that the decay length of the end modes is $\lambda = [\log (w/v)]^{-1}$, it can be made arbitrarily large by choosing values of $v$ arbitrarily closer to $w$. This, in turn, increases the effective hopping amplitude and reduces the transfer time. However, the energy gap is also drastically reduced, due to the proximity of the topological phase transition at $v=w$, making it much more susceptible to noise and disorder. In a real system with finite precision, the control parameter would have to satisfy $v < w-\delta \epsilon$ in order to keep some protection, where $\delta\epsilon$ is the scale of the combined effects of noise, disorder and experimental error\footnote{This can be easily derived from the fact that the gap in the SSH chain is $\Delta=2|w-v|$ \cite{Asboth2015}.}.

In this paper, we propose an alternative solution: the use of intermediate domain walls as amplifiers. To illustrate our approach, let us start by calculating the transfer time for a two-domain SSH chain:

\begin{align}
t_{\textrm{tr}}^{(N=2)} & =\frac{\pi\sqrt{w^{2}+v^{2}}}{\sqrt{2}(w^{2}-v^{2})}\left(\frac{w}{v}\right)^{\ell/2+1}\nonumber \\
 & =\frac{\pi\sqrt{w^{2}+v^{2}}}{\sqrt{2}(w^{2}-v^{2})}\left(\frac{w}{v}\right)^{(L+1)/4}.\label{eq:T_tr_SSH2}
\end{align}

It also scales exponentially with $L$, but the exponent is now $L/4$ instead of $L/2$, due to the presence of the intermediate $\mathcal{S}$ state. However, both results depend exponentially on $\ell /2$. This is because the effective hopping amplitude between boundary states, which ultimately determine the transfer time, depends only on the domain length $\ell$, not on $L$.

This result points to the idea that we can lose the exponential dependence altogether by fixing $\ell$ and increasing the number of domains $N$. As seen in Section \ref{sec:SDoms}, this problem is reduced to implementing a transfer in the effective model chain with $N+1$ sites, which can be achieved changing only four of the effective links, as shown in \cite{Apollaro2012,Banchi2013} for trivial 1D chains. It has also been shown that the transfer time can be made linear in the size of the effective chain, $t_{\mathrm{tr}}\sim (N+1)/J$, where $J$ is the effective hopping amplitude \cite{Banchi2015,Albanese2004,Christandl2004}. This result shows that the optimal transfer time in a family of chains of fixed $\ell$ and increasing $N$ will scale at most linearly with $N$, and therefore with the total length $L = N(\ell+1) +1$ as well.

Ultimately, our proposal works because $e^{L}\gg N e^{L/N}$ for large values of $L$ if $N$ is large enough, i.e., if sufficient domain walls are considered.

\begin{figure*}[!tph]
\mbox{%
\includegraphics[width=\textwidth]{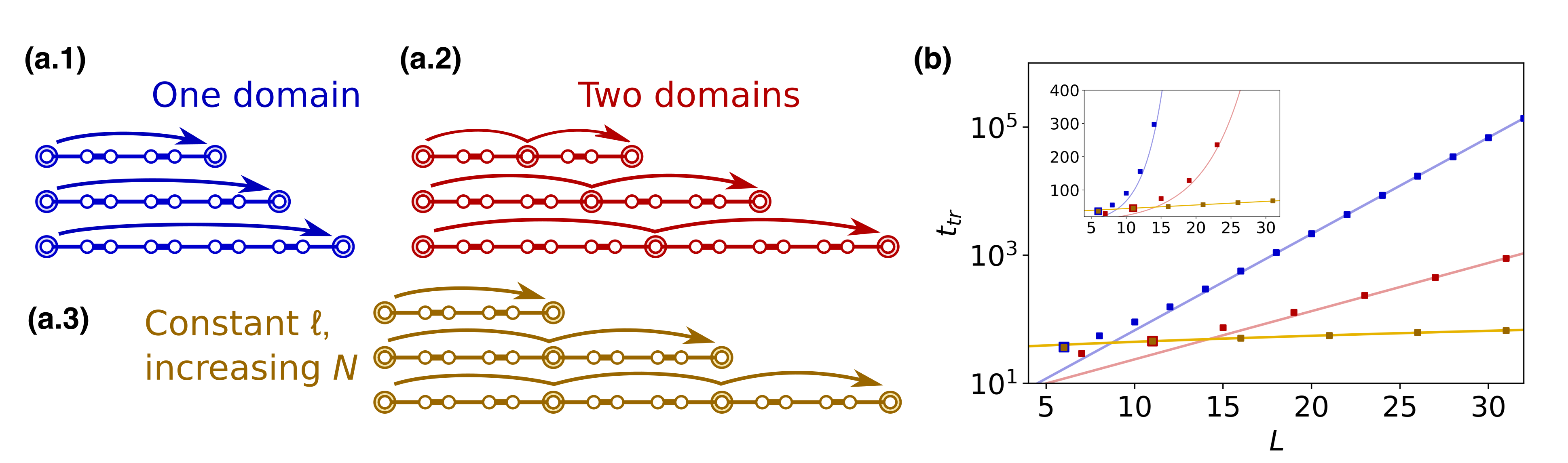}%
}\caption{Transfer time between left and right states in an SSH chain as a function
of distance. We consider (a.1) a single domain of increasing length,
in blue, (a.2) two domains of increasing length, in red, and (a.3)
an increasing number of domains of length $\ell=4$, in yellow. (b)
Transfer time $t_{\textrm{tr}}$ in the three cases as a function
of the total length of the chain $L$, color coded as in (a). A logarithmic
scale is used for the vertical axis. The first and second yellow points
coincide with a blue point and a red point, respectively. The simulations
have a maximum control parameter of $v_{\textrm{tr}}=0.5$, a preparation
time of $t_{\textrm{prep}}=15$, and a time step of $\Delta t=0.1$.
Analytical results for the transfer times {[}Eqs. (\ref{eq:T_tr_SSH1},\ref{eq:T_tr_SSH2}){]}
are included in continuous red and blue lines. A linear fit is plotted
for the yellow points. In the inset, the same data is plotted using
a linear scale, in order to appreciate the different trends of the
data. \label{fig:SSH_T_tr}}
\end{figure*}

We demonstrate this in Fig. \ref{fig:SSH_T_tr}, in which we plot $t_{\textrm{tr}}$ as a function of distance for the single- and two-domain
cases, as well as for a chain of fixed $\ell=4$ and increasing $N$.
We include both numerical results (points) and analytical estimates (lines) for one and two domains. We set $v_{\textrm{tr}}^{(1)}=0.5, t_{\textrm{prep}}=15$, and a time step of $\Delta t=0.1$ for all simulations.

The third case can be fitted to a linear function $t_{\textrm{tr}}=t_{0}+A_{0}L$, with $t_{0}=33.6$ and $A_{0}=1.08$. The first point was not included in the fit, due to the preparation time being non-negligible there. The optimal values for $v_{\textrm{tr}}^{(k\ne 1)}$ in simulations with $N>2$ were obtained numerically using the effective Hamiltonian, and are included in Table \ref{tab:Prots}, inside Appendix \ref{sec:kappas}.

The concern may arise that the addition of intermediate states could
also make the protocols more fragile against noise. However, as we
show in the following section, the topological protection of these
protocols show the expected plateau at perfect fidelity for low disorder levels
when the corresponding chiral symmetry is preserved, and when it is
not, their shortened times actually make them much more robust than
the single-domain case.

As a side note, we remark that equations (\ref{eq:T_tr_SSH1}) and (\ref{eq:T_tr_SSH2}) are only estimates because they assume a constant control parameter $v$ and do not take preparation time into account. The latter is chosen to be independent from $L$, given that the gap does not depend significantly on length \cite{Asboth2015} and is bounded above by its $L\to\infty$ value. As can be seen in the figure, the preparation time soon becomes negligible as transfer times increase.

Finally, we also want to point out that other authors have considered the notion of topological amplification in single-domain, non-Hermitian models \cite{Porras2019,Wanjura2020,Ramos2021,Wanjura2021,Ramos2022,Gomez-Leon2022}. The kind of amplification we study is similar in spirit, although its purpose and topological properties are different.

%#######################################################################################
%#######################################################################################

\subsection{Robustness against disorder \label{subsec:Sdis}}

In this Section, we use exact diagonalization to study the effects of disorder on the LR transfer protocols defined above. We use their fidelity, $f=\left|\bk{\mathcal{R}}{\psi(t_{\textrm{tr}})}\right|^{2}$, where $\ket{\psi(t_{\textrm{tr}})}$
is the state obtained at the end of the protocol.

We consider the single-domain and the four-domain case. In order to benchmark their performance, we also include a simple non-topological protocol in a 1D chain with hopping amplitude $w=1$, driven by tuning the chemical potential at the end sites of the chain. It starts at a low value $-\mu_0$ in order to create a potential well at either end, and is then adiabatically increased up to zero during a time $t_{\textrm{prep}}$. The two wells are then re-established adiabatically, by lowering the potential again. We consider $\mu_0 = 10$ and $t_{\textrm{prep}}=t_{\textrm{tr}}/2$, and then find the optimal transfer time in the ideal system that satisfies $f>0.985$\footnote{Greater fidelities can be obtained by increasing $\mu_0$, but the value was kept at $10w$ in order to keep a consistent energy scale.}.

We consider quenched disorder in the time scale of the protocols (i.e.
constant in time for each realization). We note that we do not calibrate the transfer time for each realization as done in \cite{Lang2017}, given
that this might not be feasible in a real system. We study two
disorder regimes: off-diagonal disorder, which preserves chiral
symmetry, and a case with both diagonal and off-diagonal
disorder, which breaks it. The symmetry-preserving disorder is modeled as:

\begin{align}
-v_j^{\textrm{dis}} =-|v+\delta J R_{j}| \\
-w_j^{\textrm{dis}} =-|w+\delta J R_{j}|
\label{eq:SSHoffDdis}
\end{align}

where $\delta J$ is the level of off-diagonal disorder, $-v_j^{\textrm{dis}}$
and $-w_j^{\textrm{dis}}$ are the hopping amplitudes in the disordered model, where the amplitude with site index $j$ connects sites $j$ and $j+1$. $R_{j}$ are random numbers in the interval $[-0.5,0.5]$, sampled uniformly. It is important to note that the disorder at each link is independent from all the others, given that each link has a different $j$. Fluctuations in the control parameters $v$ do not break topological protection, something crucial for the robustness of the protocol.

For the case with general disorder (both diagonal and off-diagonal), we consider:

\begin{equation}
\mu_{j} = \delta\mu R_{j}^{(\mu)}\label{eq:SSHmuDis}
\end{equation}
\begin{align}
-v_j^{\textrm{dis}} =-|v+\delta J R_{j}^{(J)}| \\
-w_j^{\textrm{dis}} =-|w+\delta J R_{j}^{(J)}|,
\label{eq:SSHoffDdis}
\end{align}
where $\mu_{j}$ is the chemical potential (i.e. on-site energy) in site $j$, $\delta\mu$ and $\delta J$ represent the level of diagonal
and off-diagonal disorder respectively, and all different $R_j$ variables
are independent random numbers in the interval $[-0.5,0.5]$, and
uncorrelated for all different sites and bonds.

\begin{figure}[!tph]
\mbox{%
\includegraphics[width=1\columnwidth]{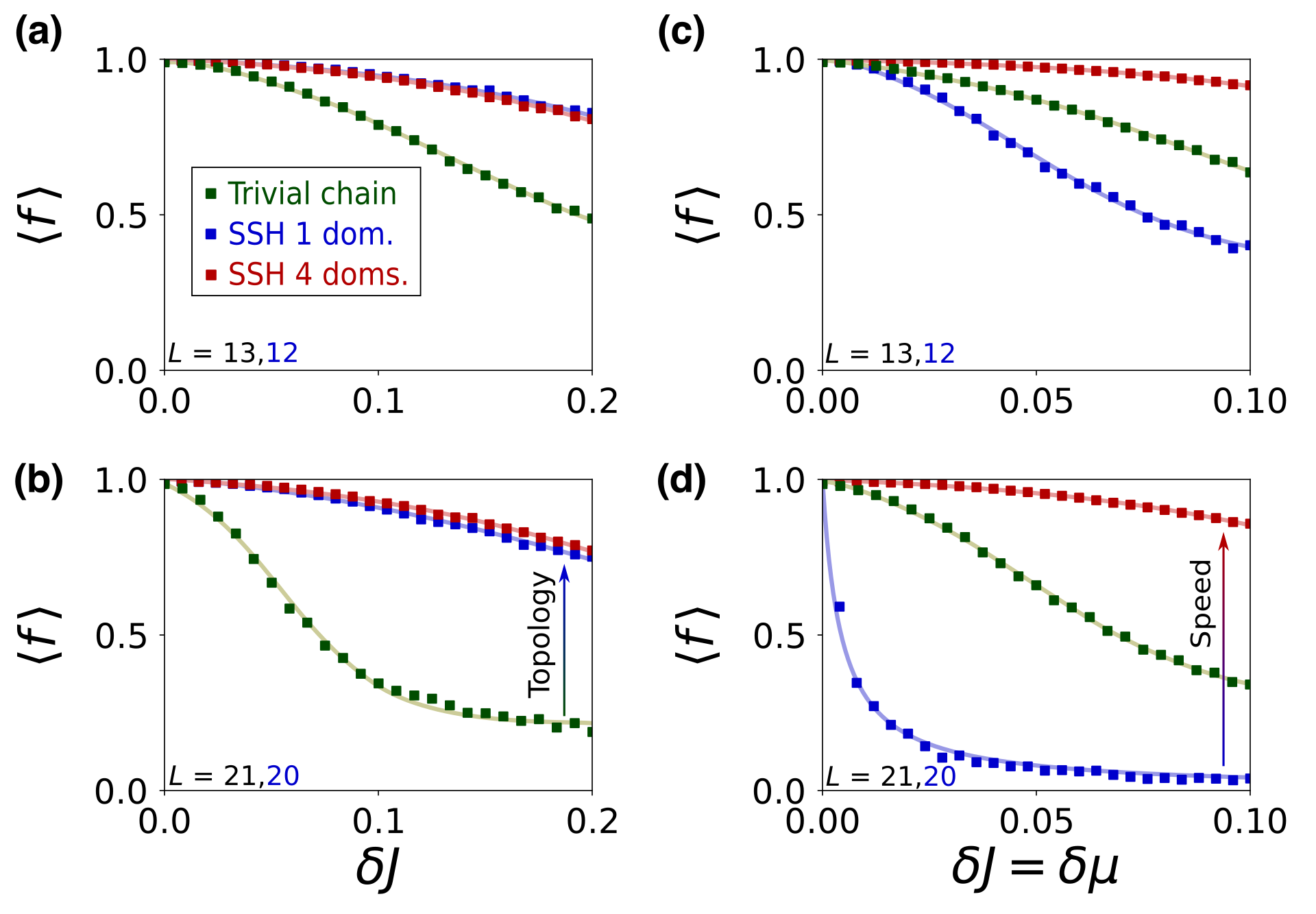}%
}\caption{Fidelity of the SSH chain transfer protocols against quenched disorder.
We compare a transfer protocol in a trivial chain to the single- and
four-domain topological SSH chain protocols, for a fixed length $L$.
We study the cases $L=13,21$, except in the single-domain SSH chain,
where the total length has to be even, so we choose $L=12,20$. We
consider off-diagonal disorder (a,b) and diagonal and off-diagonal
disorder at the same time (c,d), see main text. The effect of topological
protection can be seen in (a,b), where the topological protocols show
a plateau for low values of disorder. Note the different horizontal
scale for (a,b) and (c,d). In the case with symmetry-breaking disorder,
the multidomain protocol performs much better than the single-domain
one, due to its shorter transfer time. Sigmoid functions fitted
to the data are provided as a guide for the eye. Transfer times for
$L=13,12$ are $t_{\textrm{tr}}=332.3,156.5,35.0$ for the trivial
chain and the single- and four-domain SSH chain, respectively. In
the same order, the times for $L=21,20$ are $t_{\textrm{tr}}=800.5,2175.4,55.9$.
\label{fig:Dis_SSH}}
\end{figure}

The results can be seen in Fig. \ref{fig:Dis_SSH} for systems of length 13 (a,c) and 21 (b,d)---or, in the case of the single-domain SSH chain, which always has an even number of sites, $L=12$ and $20$. We show the average over 1000 realizations As expected, the topological protocols show a plateau in their fidelity results for off-diagonal disorder, which persists up to more than $0.2w$. In the case of general disorder (notice the different horizontal scale), the poor performance of the single-domain SSH chain in the $L=20$, which is worse than the trivial case, illustrates the problem of exponentially long times: many more errors can add up, losing the advantage in the presence of even small amounts of symmetry-breaking disorder. In contrast, the four-domain chain shows a much better performance up to more than $0.1w$, showing a clear advantage over the single-domain and the trivial one, due to its much shorter transfer time.

We include a more detailed analysis of the different factors that we
observed can affect the robustness of the studied protocols in
Appendix \ref{sec:Disorder_Ap}.

The phase acquired in the transfer is much more robust in the topological case than in the trivial one, due to it being a geometrical phase instead of a dynamical one. This is an important point when considering the quantum information applications of these protocols, including coupling to external qubits \cite{Lang2017} and native implementations of the model in quantum processors \cite{Mei2018}. We elaborate on this point in Section \ref{sec:TwoStateTr}, and showcase its usefulness in a minimal case involving the transfer of two superposed states in Section \ref{sec:TwoStateTr}.

%#######################################################################################
%#######################################################################################

\section{The Creutz ladder: walls with two states \label{sec:Creutz}}

We now consider another topological insulator: the imbalanced Creutz ladder (ICL). Its main difference with the SSH chain is the presence of flat bands in the balanced limit, and the existence of domain walls with two protected states, which had not been studied before in the literature. These two phenomena can be used to implement more complex transfer protocols than in the SSH chain, in which the exponential speed-up is also present.

\subsection{Topological phases and domain walls \label{subsec:TopoDomW}}%Add Eff Ham

\begin{figure}[!tph]
\mbox{%
\includegraphics[width=1\columnwidth]{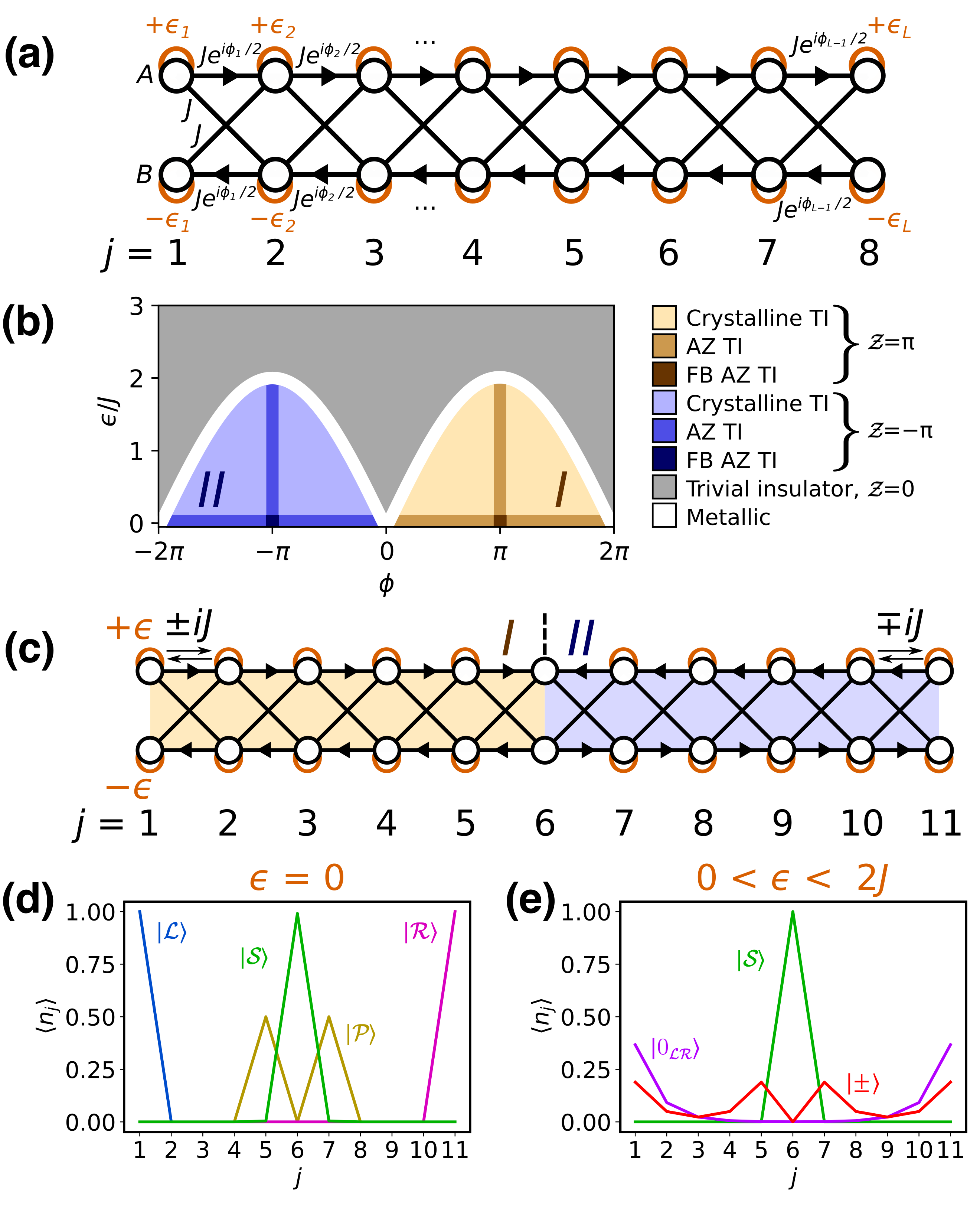}%
}\caption{(a) Imbalanced Creutz ladder in the general case. The energy
imbalance parameters $\epsilon_{j}$ and the magnetic Peierls phases
$\phi_{j}$ determine the topology, and can be used to define different
domains. All hopping amplitudes have an additional minus sign. (b)
Topological phase diagram of the ICL. It is divided
into two topological phases with Zak phases $\sZ=\pm\pi$, and
a trivial phase with $\sZ=0$. The topology of the crystalline TI
regions cannot be described by the standard Altland-Zirnbauer symmetry
classes \cite{Altland1997}. Those which can are marked with ``AZ''.
The flat-band points are labeled ``FB''. (c) Two-domain Creutz ladder
with no imbalance in its domain wall ($\epsilon_{6}=0$), and with
an energy imbalance parameter of $\epsilon$ in all other rungs.
This model has four topological states: the left and right end modes,
and two states at the domain wall. (d) Spatial distribution of the
AB-caged topological states in a two-domain Creutz ladder with $\epsilon=0$:
left, right and $\mathcal{S}$- and $\mathcal{P}$-type states. All
of them are compact and pinned at zero energy. (e) Spatial distribution
of the topological states in the system in (c), for $\epsilon=J$.
A pair of antibonding and bonding states $\ket{\pm}$ appear with
small, opposite energies, while the other two states remain at zero
energy. \label{fig:Creutz1}}
\end{figure}

We consider an imbalanced Creutz ladder Hamiltonian:

\begin{align}
\mathcal{H} & =-\sum_{j=1}^{L-1}\sum_{\sigma=A,B}\!\!\left[J\xi_{j,\sigma}c_{j+1,\sigma}^{\dagger}c_{j,\sigma}+Jc_{j+1,\overline{\sigma}}^{\dagger}c_{j,\sigma}+h.c.\right]\nonumber \\
 & +\sum_{j=1}^{L}\sum_{\sigma=A,B}s_{\sigma}\epsilon_{j}c_{j,\sigma}^{\dagger}c_{j,\sigma}
\end{align}

where $j=1,\dots,L$ labels the different rungs, $\sigma=A,B$ designates
the two legs, with $\overline{A}=B$ and vice versa, $\xi_{j,\text{\ensuremath{\sigma}}}=e^{is_{\sigma}\phi_{j}/2}$,
with $s_{\sigma}=\delta_{\sigma,A}-\delta_{\sigma,B}$, $\phi_{j}$
is the magnetic flux in the $j$-th plaquette, $J$ is the horizontal
and diagonal hopping amplitude and $2\epsilon_{j}$ is the energy
imbalance between the two legs in the $j$-th rung {[}see Fig. \ref{fig:Creutz1}
(a){]}.

In a ladder with $\epsilon_{j}=\epsilon$ and $\phi_{j}=\phi$ $\forall j$, the topology of the system depends on $\phi$ and $\epsilon/J$, as can be seen in Fig. \ref{fig:Creutz1} (b). The system has two distinct nontrivial phases when $\epsilon<2J$, with Zak phases $\sZ=\pm \pi$. When $\phi=\pm\pi$, this corresponds to winding numbers $\nu=\pm1$. There is a single trivial phase ($\sZ=0$) when $\epsilon>2J|\sin (\phi/2)|$ \cite{Li2015,Sun2017,Junemann2017}. When in a topological phase, zero modes appear at the ends of the ladder.

Furthermore, the wall between domains that belong to different phases,
with winding numbers $\nu_{1}$ and $\nu_{2}$, will hold $|\Delta\nu|$
zero modes, where $\Delta\nu=\nu_{2}-\nu_{1}$. Due to the particular
form of the chiral symmetry, these states will have a positive chirality
if $\Delta\nu$ is \emph{negative}, and vice versa. As seen above, a topological-to-trivial
wall will support one zero mode, while the wall separating phases
with $\nu=\pm1$ will support two zero modes with the same chirality.

%#######################################################################################
%#######################################################################################

We consider a Creutz ladder with $N$ topological domains, each with an alternating winding number of $\nu=\pm 1$. We set $J=1$. Each domain has $\ell \geq 2$ inner rungs, and the length of the ladder is $L=N(\ell+1)+1$ (the number of sites is $2L$). This model is obtained with the following parameter values:

\begin{equation}
\phi_{j}=\begin{cases}
\pi & \textrm{if }1\leq j\mod2(\ell+1)\leq\ell+1\\
-\text{\ensuremath{\pi}} & \textrm{otherwise }
\end{cases}
\end{equation}

\begin{equation}
\epsilon_{j}=\begin{cases}
\epsilon_{w}^{(D[j])} & \textrm{if }j\mod\ell+1=1\textrm{ (domain walls)}\\
\epsilon_{b}^{(D[j])} & \textrm{otherwise (bulk sites)},
\end{cases}
\end{equation}

where $D[j]=\lceil(j-1)/(\ell+1)\rceil$ is the domain number to which
rung $j$ belongs.
For this purpose, we label each wall just like the domain to its left, leaving the leftmost rung of the ladder as belonging to domain $D[j=1]=0$. An example with $\ell=4$ and two domains can be seen in Fig \ref{fig:Creutz1} (c).

If we set $\epsilon_{j}=0\,\,\forall j$ (balanced case) with the flux values defined above, the model will have left and right topological end states, as well as two topological states, $\ket{\mathcal{S}_{k}}$ and $\ket{\mathcal{P}_{k}}$, in each domain wall $k=1,\ldots,N-1$. Given that the system presents AB caging \cite{Junemann2017,Zurita2019}, all of them will be compact,
i.e. localized in only a few nearby sites. An $N$-domain CL defined
in this way has $2N$ compact topological states pinned at exactly
zero energy: $(N-1)$ $\mathcal{S}$- and $\mathcal{P}$-type states,
and the left and right end modes. All of them are localized in two
sites, except for the $\mathcal{P}$ states, which are localized
in four. Their form is:

\begin{align}
\ket{\mathcal{L}} & =\frac{\ket{1,A}-i\ket{1,B}}{\sqrt{2}}\label{eq:L}\\
\ket{\mathcal{R}} & =\frac{\ket{L,A}+(-1)^{N+1}i\ket{L,B}}{\sqrt{2}}\label{eq:R}\\
\ket{\mathcal{S}_{k}} & =\frac{\ket{j_{k},A}+(-1)^{k+1}i\ket{j_{k},B}}{\sqrt{2}}\label{eq:M1-1}\\
\ket{\mathcal{P}_{k}} & =\frac{1}{2}[\ket{j_{k}-1,A}-\ket{j_{k}+1,A}+\nonumber \\
 & (-1)^{k+1}i(\ket{j_{k}-1,B}-\ket{j_{k}+1,B})],\label{eq:P}
\end{align}

where $k=1,\ldots,N-1$, and $j_{k}=k(\ell+1)+1$ is the rung in which
the $k$-th wall is located. We use $\mathcal{S}$ and $\mathcal{P}$
as nomenclature in an analogy with the $s$ and $p$ orbitals in an atom, with which they share symmetry properties. We show a two-domain Creutz ladder in Fig. \ref{fig:Creutz1} (c), and the four protected states it has when balanced in (d), using their rung occupation number:

\begin{equation}
\langle n_{j}\rangle=\langle n_{j,A}\rangle+\langle n_{j,B}\rangle.
\end{equation}

We group left, right and $\mathcal{S}$-type states under the name of ``computational states'' for convenience.

We discuss how to prepare these compact states in a topologically
protected way starting from a particle in a single site in Appendix
\ref{sec:CSSH}.

If the imbalance is now switched on for all sites except those in a domain wall, $\epsilon_{w}^{(0)}=\epsilon_{w}^{(N)}=\epsilon_{b}^{(D)}=\epsilon<2J\,\,\forall D$,
$\epsilon_{w}^{(D)}=0\,\,\forall D\in[1,N-1]$; then, the left, right and $\mathcal{P}$-type topological states will acquire an exponential profile, overlap with each other and hybridize. Their analytical expression before hybridization is:
\begin{widetext}
\begin{align}
\ket{\mathcal{L}} & =-i\mathcal{N_{L}}\sum_{j=1}^{\ell+1}\left(\frac{2J}{i\epsilon}\right)^{-j}\ket{j}\otimes\left(\begin{array}{c}
1\\
-i
\end{array}\right)\label{eq:Lm-1-1}\\
\ket{\mathcal{R}} & =(-1)^{N+1}i\mathcal{N_{R}}\sum_{j=L-\ell}^{L}\left((-1)^{N}\frac{2J}{i\epsilon}\right)^{j-L-1}\ket{j}\otimes\left(\begin{array}{c}
1\\
(-1)^{N+1}i
\end{array}\right)\label{Rm-1-1}\\
\ket{\mathcal{P}_{k}} & =(-1)^{k+1}i\mathcal{N}_{\mathcal{P}_{k}}\left[\sum_{j=(k-1)(\ell+1)+2}^{k(\ell+1)}\left((-1)^{k}\frac{2J}{i\epsilon_l}\right)^{j-k(\ell+1)-1}\ket{j}\otimes\left(\begin{array}{c}
1\\
(-1)^{k+1}i
\end{array}\right)\right.\nonumber \\
 & \left.-\sum_{j=k(\ell+1)+2}^{(k+1)(\ell+1)}\left((-1)^{k}\frac{2J}{i\epsilon_r}\right)^{k(\ell+1)-j+1}\ket{j}\otimes\left(\begin{array}{c}
1\\
(-1)^{k+1}i
\end{array}\right)\right],\label{eq:Pm-1-1}
\end{align}
\end{widetext}

where $\mathcal{N}_{\mathcal{L},\mathcal{R},\mathcal{P}}$ are normalization
constants, and can be approximated for $\ell\gtrsim 3$ as:

\begin{equation}
\mathcal{N_{L}}=\mathcal{N_{R}}\approx\sqrt{\frac{4J^{2}/\epsilon^{2}-1}{2}}
\end{equation}
\begin{equation}
\mathcal{N_{P}\approx}\frac{1}{\sqrt{2}}\left(\frac{1}{4J^{2}/\epsilon_{l}^{2}-1}+\frac{1}{4J^{2}/\epsilon_{r}^{2}-1}\right)^{-1/2},
\end{equation}

where $2\epsilon_{l,r}$ are the values of the energy imbalance in
the domains to the left and right of the relevant domain wall. Every state is confined to its adjacent domains. The form of the two end modes was also derived in \cite{Cobanera2017,Cobanera2018} for a runged CL. Note that the $\mathcal{P}$ states become asymmetrical if $\epsilon_l\ne\epsilon_r$. The
gauge was chosen to coincide with Eqs. (\ref{eq:L}-\ref{eq:P}).
We include the derivation of these states in Appendix \ref{sec:DerivStates}\footnote{Like in the SSH chain, finite size effects can make the states deviate slightly from these expressions for small domains close to the phase transition ($\delta_{\mathcal{L}}=\norm{\ket{\mathcal{L}}_{\mathrm{num}}-\ket{\mathcal{L}}_{\mathrm{analyt}}}=0.16$
for $\ell=2$ and $\epsilon=1.5J$), but they are highly accurate in general ($\delta_{\mathcal{L}}=0.016$ for $\ell=4$ and $\epsilon=J$).}.

Given that the potential at all domain wall sites is set to zero, the magnetic interference keeping the $\mathcal{S}$ states localized on two sites is not broken, and each of these states remains compact and isolated from the rest of the system. This is illustrated in Fig. \ref{fig:Creutz1} (e) for a two-domain ladder, where bonding and antibonding states appear, with the form $\ket{\pm}=(\ket{\mathcal{L}}-\ket{\mathcal{R}})/2\pm i^{\ell}\ket{\mathcal{P}}/\sqrt{2}$,
as well as a dark state at zero energy, $\ket{0_{\mathcal{LR}}}=(\ket{\mathcal{L}}+\ket{\mathcal{R}})/\sqrt{2}$. States $\ket{\pm}$, while having the same total occupation at each rung, have a higher occupation in one of the legs of the ladder, and their positive and negative energies can be explained by the imbalance.

\subsection{Effective Hamiltonian}

The effective Hamiltonian for the protected states will now include $N-1$ isolated $\mathcal{S}$ states, while all the others will hybridize as in the SSH chain case, forming an effective 1D chain with $N+1$ sites:

\begin{align}
\mathcal{H}_{\mathrm{eff}}=v_{1}\ketbra{\mathcal{P}_{1}}{\mathcal{L}}+v_{N}\ketbra{\mathcal{R}}{\mathcal{\mathcal{\mathcal{P}}}_{N-1}}+\nonumber\\
+\sum_{k=2}^{N-1}v_{k}\ketbra{\mathcal{\mathcal{\mathcal{P}}}_{k}}{\mathcal{\mathcal{\mathcal{P}}}_{k-1}}+h.c.\label{eq:Heff}
\end{align}

Using expressions (\ref{eq:Lm-1-1}-\ref{eq:Pm-1-1}), we obtain the
following expression for the hopping amplitudes:

\begin{equation}
v_{k}=2(-1)^{d+p_{k-1}}\epsilon\mathcal{N}_{k}\mathcal{N}_{k-1}\left[(-1)^{k}\frac{2iJ}{\epsilon}\right]^{-d-2},\label{eq:v_k}
\end{equation}

where $\mathcal{N}_{k},\mathcal{N}_{k-1}$ are the normalization constants
for the involved states and $d$ is the distance between the maxima
of the $k$-th and the $(k-1)$-th topological states, with $\ket{\mathcal{L}}$
being the zeroth state and $\ket{\mathcal{R}}$ being the $N$-th. The distance is $d=\ell-1$ between two $\mathcal{P}$ states, $d=\ell$ between an end state and a $\mathcal{P}$ state, and $d=\ell+1$ between
the two end states in a single-domain ladder. $p_{k-1}=0$ if the $(k-1)$-th state is a $\mathcal{P}$ state, and $p_{k-1}=1$ otherwise.

We can use this effective Hamiltonian to understand the dynamics
of the topological states. If the system is initialized in state $\ket{\mathcal{L}}$,
time evolution will eventually take it to state $\ket{\mathcal{R}}$,
using each $\mathcal{P}$-type state of the ladder as a way to leapfrog
its domain wall without affecting its $\mathcal{S}$-type state. Now, the energy imbalance in each domain can be used as a control parameter to implement a transfer protocol between the two end states.

%#######################################################################################
\begin{figure*}[!tph]
\mbox{%
\includegraphics[width=1\textwidth]{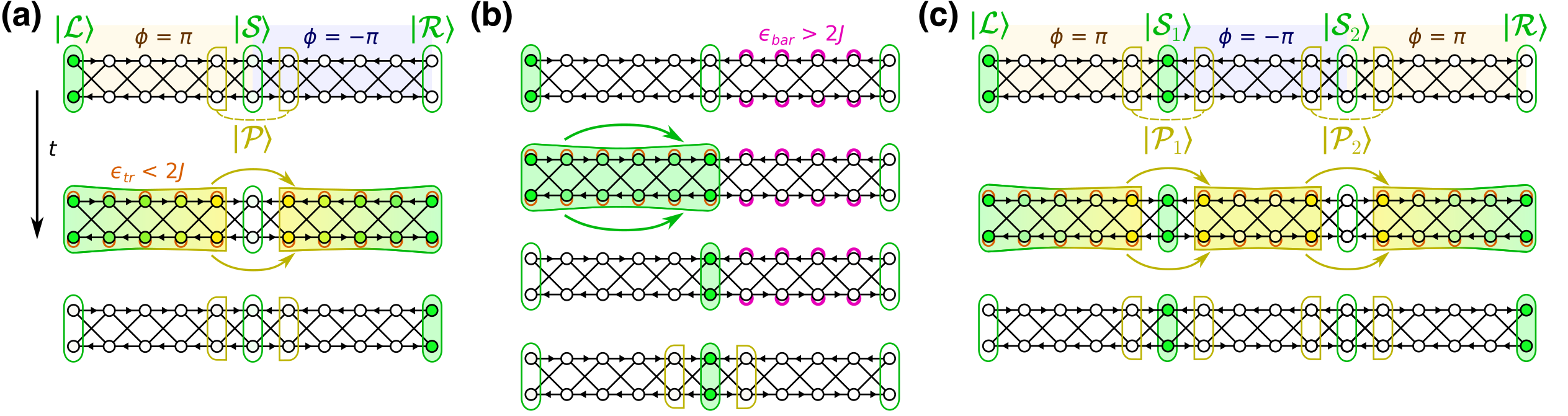}%
}\caption{(a) $\mathcal{L}$-to-$\mathcal{R}$ transfer protocol in a two-domain
Creutz ladder with $\ell=4$, induced by an energy imbalance of $\pm\epsilon_{\mathrm{tr}}$.
One of the topological states in the domain wall, $\ket{\mathcal{S}}$,
remains pinned at zero energy due to AB caging, while the other one,
$\ket{\mathcal{P}}$, hybridizes with the left and right states to
allow the transfer. (b) $\mathcal{L}$-to-$\mathcal{S}$ transfer
in a two-domain ladder. The initial state, not shown, is the same
as in (a). The second domain has to be taken to the trivial phase
using a control parameter of $\epsilon_{\mathrm{bar}}>2J$ before
inducing the transfer process. (c) $\mathcal{L}$-to-$\mathcal{R}$
transfer in a three-domain ladder, starting in state $(\ket{\mathcal{L}}+\ket{\mathcal{S}_{1}})/\sqrt{2}$.
The final state is $(\ket{\mathcal{S}_{1}}-\ket{\mathcal{R}})/\sqrt{2}$.
As can be seen, the transfer does not affect the intermediate $\mathcal{S}$-type
states. \label{fig:transf}}
\end{figure*}
%#######################################################################################

\subsection{More complex transfer protocols}%Add S states and phases

An LR transfer can be achieved using the same type of pulses as in the SSH chain, now tuning the energy imbalance parameters for each domain $D$:

\begin{align}
&\epsilon_{\mathrm{tr}}^{(D)}(t)=\nonumber\\
&=
\begin{cases}
\epsilon_{\mathrm{tr}}^{(D)}\sin^{2}(\Omega t) & \textrm{for }0 \leq t<t_{\mathrm{prep}}\\
\epsilon_{\mathrm{tr}}^{(D)} & \textrm{for }t_{\mathrm{prep}} \leq t<t_{\mathrm{tr}}-t_{\mathrm{prep}}\\
\epsilon_{\mathrm{tr}}^{(D)}\sin^{2}[\Omega(t-t_{\mathrm{tr}})] &  \textrm{for }t_{\mathrm{tr}}-t_{\mathrm{prep}}\leq t\leq t_{\mathrm{tr}}.
\end{cases}\label{eq:CParam-1}
\end{align}

In addition to LR transfers, we are going to consider other possibilities: left-to-$\mathcal{S}$, $\mathcal{S}$-to-$\mathcal{S}$ and $\mathcal{S}$-to-right transfers. We illustrate some possibilities in Fig. \ref{fig:transf}. In subfigure (a), a two-domain LR transfer is illustrated in the ICL. Notice how the $\mathcal{S}$ state is left invariant, because we do not apply an imbalance in its domain wall.

This is the motivation to consider other protocols: if some amplitude had been transferred into that $\mathcal{S}$ state before the LR transfer, it would have survived that second transfer. This allows us to couple any two computational states without disturbing the rest, providing all-to-all connectivity, similar to a chain of T-junctions \cite{Ezawa2020}, but in a homogeneous quasi-1D lattice.

In Fig. \ref{fig:transf} (b), we show how to implement a left-to-$\mathcal{S}$ transfer. Firstly, the $\mathcal{P}$ state must disappear in order to not get mixed with the $\mathcal{S}$ state, and we need a way for the transfer to stop at the intended wall. Both of these problems are solved either by disconnecting the rest of the ladder or---less drastically---by adiabatically setting up an energy imbalance of $\epsilon_{\textrm{bar}}>2J$ in the next domain, which will act as a barrier inhibiting the effective hopping amplitude across its domain, thereby stopping the transfer. The reason is that this makes the domain trivial, turning the wall into a topological-to-trivial wall, only holding the $\mathcal{S}$ state, which remains unperturbed during this process. A large value of the barrier, $\epsilon_{\textrm{bar}}\gtrsim 20 \epsilon_{\mathrm{tr}}$, has been observed to be more effective in this task.

%#######################################################################################
\begin{figure}[!tph]
\mbox{%
\includegraphics[width=1\columnwidth,trim={0 0 0 10},clip]{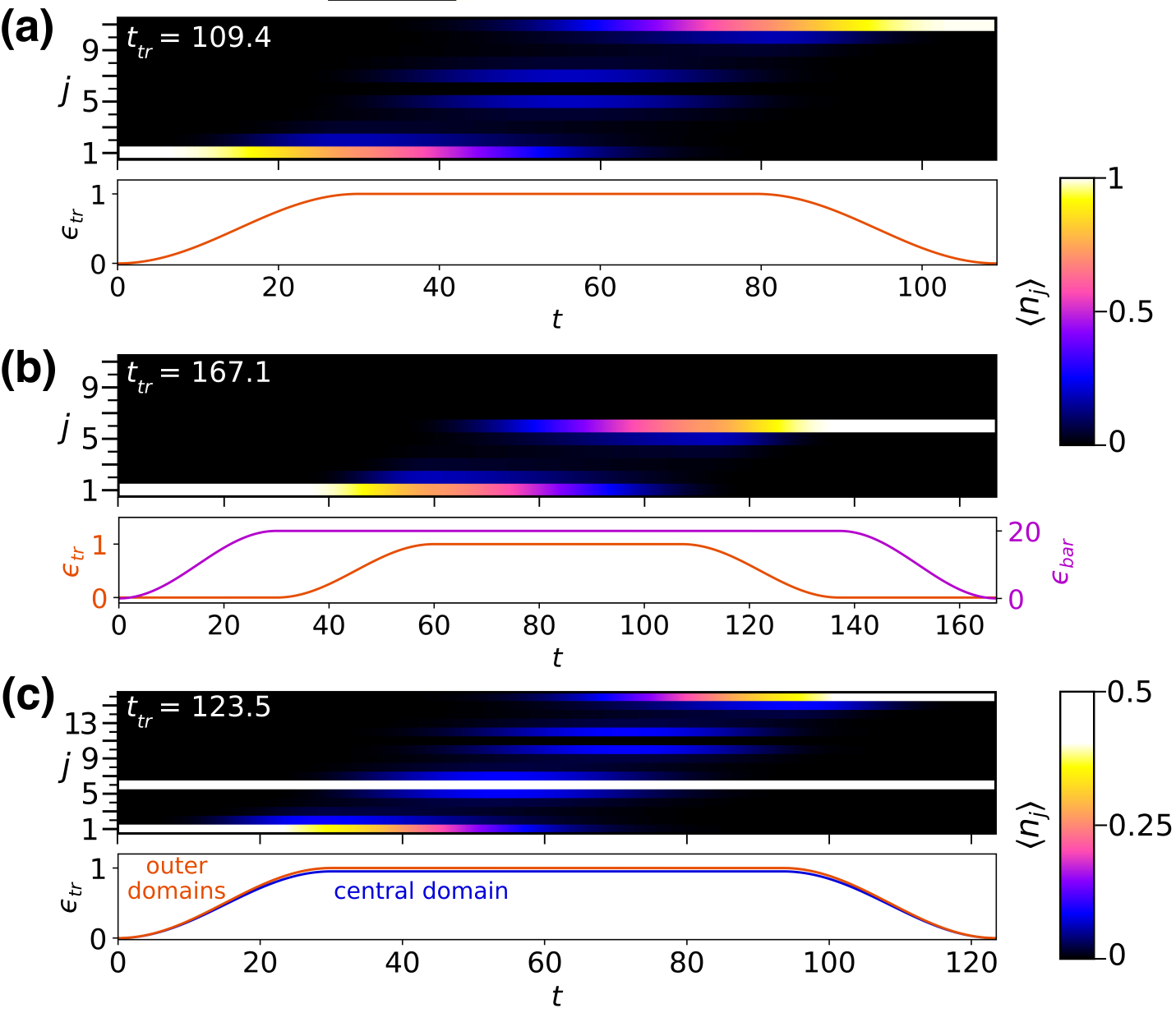}%
}\caption{Exact diagonalization simulations of the processes in Fig. \ref{fig:transf},
and the control pulses used. (a) Left-to-right transfer
in a two-domain ladder. (b) Left-to-center transfer in a two-domain
ladder. Note the different scales for the transfer-inducing parameter $\epsilon_{\textrm{tr}}$
and the barrier $\epsilon_{\textrm{bar}}$. (c) Left-to-right transfer in a three-domain ladder, acting
on the initial state $(\ket{\mathcal{L}}+\ket{\mathcal{S}_{1}})/\sqrt{2}$.
The component at $\mathcal{\ket{\mathcal{L}}}$ gets transferred,
while the component at $\ket{\mathcal{S}_{1}}$ remains unperturbed.
The control parameter in the central domain takes a maximum value
of $0.952$, see main text. The rung occupation number is represented
as a function of time $t$ for each rung $j$. Note the different
colormaps used for subfigures (a,b) and subfigure (c). Preparation
times of $t_{\textrm{prep}}=t_{\textrm{prep}}^{\prime}=30$ are
used. All times are expressed in units of $\hbar/J$. \label{fig:sims}}
\end{figure}
%#######################################################################################

Then, the usual control pulse can be used (now including the sites in the wall) to transfer the left state to the $\mathcal{S}$ state, which acts now like the right end of a topological ladder. After the transfer is done, the barrier can be adiabatically switched off. These protocols can also reach over several walls, exactly in the same way as LR transfers, without disturbing any other $\mathcal{S}$ states in their way.

During the transfer, the $\mathcal{S}$ takes an exponential profile into the domain to its left (or right, if the barrier was set up to its left):
\begin{widetext}
\begin{align}
\ket{\mathcal{S}_{k}}_{\mathrm{left}} & =\mathcal{N}_{\mathcal{S}_{k}}\sum_{j=(k-1)(\ell+1)+2}^{k(\ell+1)+1}\left((-1)^{k}\frac{2J}{i\epsilon}\right)^{j-k(\ell+1)-1}\ket{j}\!\otimes\left(\begin{array}{c}
1\\
(-1)^{k+1}i
\end{array}\right)\\
\ket{\mathcal{S}_{k}}_{\mathrm{right}} & =\mathcal{N}_{\mathcal{S}_{k}}\sum_{j=k(\ell+1)+1}^{(k+1)(\ell+1)}\left((-1)^{k}\frac{2J}{i\epsilon}\right)^{k(\ell+1)-j+1}\ket{j}\!\otimes\left(\begin{array}{c}
1\\
(-1)^{k+1}i
\end{array}\right),
\end{align}
\end{widetext}
where the subindex indicates the side towards which the state extends and $\epsilon$ is the imbalance parameter in the relevant domain.

Just as in the SSH case, any protocol that spans over three domains or more, like the one depicted in Fig. \ref{fig:transf} (c), will need control pulses of different height for each pair of domains related by spatial inversion.

In Fig. \ref{fig:sims}, we show exact diagonalization results for the rung occupation as a function of time for each of the protocols shown in Fig. \ref{fig:transf}, along with the control pulses used. Note the different vertical scales for $\epsilon_{\textrm{tr}}$ and $\epsilon_{\textrm{bar}}$ in (b), and the fact that, in (c), the part of the wavefunction stored in state $\mathcal{S}_1$ is invariant throughout the protocol. A maximum control parameter of $\epsilon_{\textrm{tr}}=J=1$ is chosen, halfway to the topological phase transition at $2J$, which makes the decay length $\lambda = 3.32$. For this reason, we choose domains of length $\ell=4$. The preparation time for $\epsilon_{\textrm{tr}}$ and for $\epsilon_{\textrm{bar}}$ are equal, $t_{\textrm{prep}}=t_{\textrm{prep}}^\prime=30$.

\begin{figure}[!tph]
\mbox{%
\includegraphics[width=1\columnwidth]{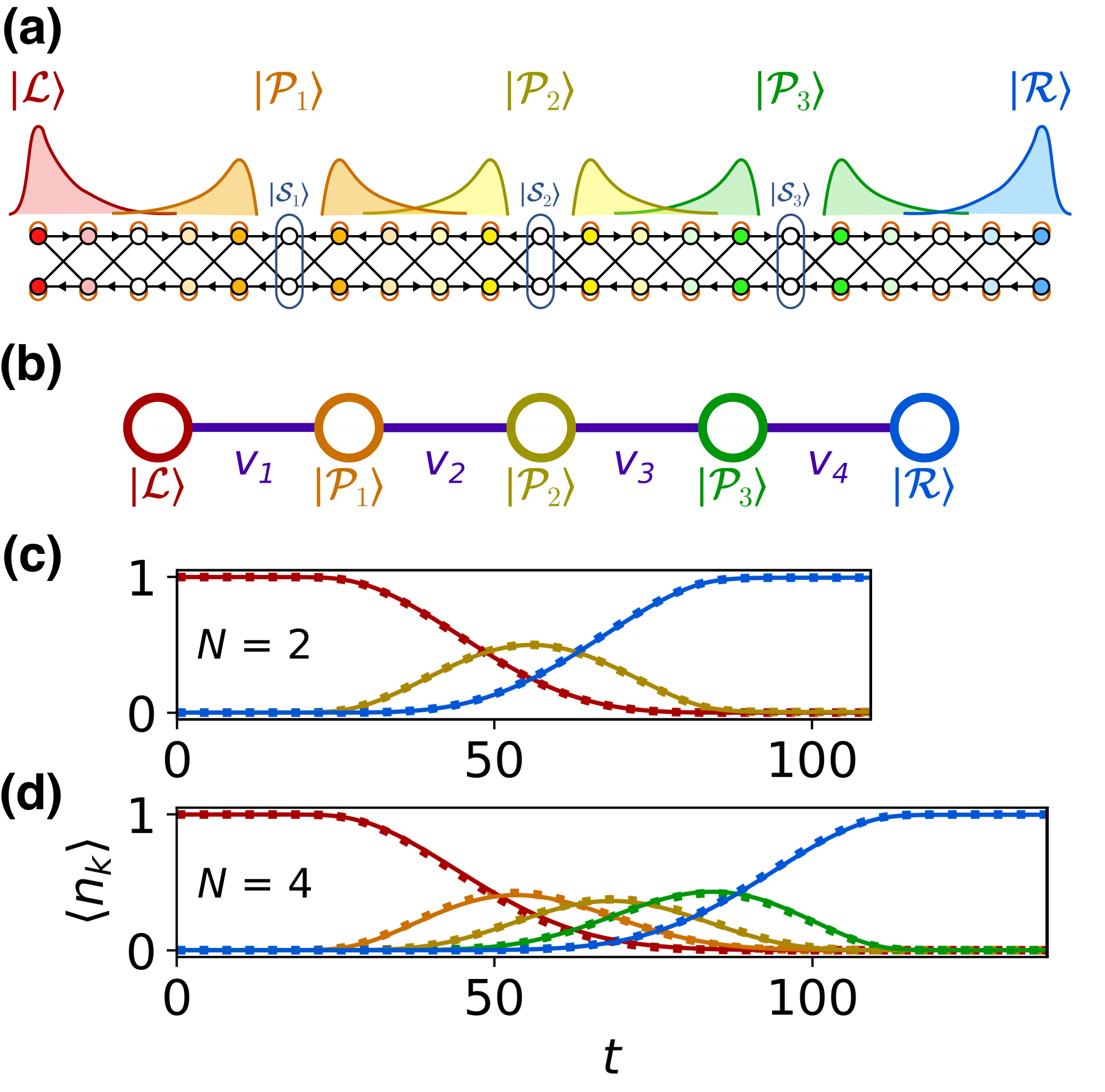}%
}\caption{(a) Topological states in a four-domain
Creutz ladder. (b) Associated effective model, corresponding to a
1D chain. (c) Occupation of the topological states during an LR transfer
in a two-domain ladder with $\ell=4$ ($L=11$), $\epsilon_{\textrm{tr}}=1$ and $t_\textrm{prep}=30$, where the only $\mathcal{P}$
state is represented in yellow. Numerical results for the full Creutz
ladder Hamiltonian are shown in continuous line, while the analytical
prediction by the effective Hamiltonian is plotted in dotted lines.
This process is also shown in Fig. \ref{fig:sims} (a). (d) Occupation
of the topological states in an LR transfer in the four-domain ladder
depicted in (a), with $\ell=4$ ($L=21$), $\epsilon_{\textrm{tr}}=1$ and $t_\textrm{prep}=30$. \label{fig:Eff}}
\end{figure}

Finally, to demonstrate the accuracy of the effective model,
transfers between the left and right end modes were simulated in two-
and four-domain ladders [Fig. \ref{fig:Eff} (a)] with $\ell=4$, and then compared to the results
predicted by the effective model [Fig. \ref{fig:Eff} (b)]. The topological state occupation
$\langle n_{k}\rangle$ is shown in subfigures (c,d) at each time
$t$, with $\langle n_{0}\rangle=|\bk{\mathcal{L}}{\psi(t)}$$|^{2}$,
$\langle n_{k}\rangle=|\bk{\mathcal{P}_{k}}{\psi(t)}|^{2}$ for $1<k<N-1$,
and $\langle n_{N}\rangle=|\bk{\mathcal{R}}{\psi(t)}|^{2}$. As can
be seen, both results are almost identical. We use $\epsilon_{\textrm{tr}}=J=1$, $t_{\textrm{prep}}=30$.

The phase acquired by a component of the wavefunction in the transfer is now given by:

\begin{equation}
\zeta(\ell,n_{w},x,\varsigma)\!=\!\begin{cases}
\!(-1)^{n_{w}/2+\delta_{x,\varsigma}}[(-1)^{\delta_{x,\varsigma}}i]^{\ell} & \!\!\textrm{for even } \! n_{w}\\
\!(-1)^{(n_{w}-1)/2} & \!\!\textrm{for odd } \! n_{w},
\end{cases}\label{eq:zetaCL}
\end{equation}

where $n_{w}$ is the number of domain walls between the transferred states, $x=\pm1$ is the chirality of the leftmost transferred state (e.g., $x=-1$ if state $\ket{\mathcal{L}}$ is involved), and $\varsigma=1\,(-1)$ indicates the direction of transfer, from left to right (from right to left). The two latter quantities are then compared in a Kronecker delta
$\delta_{x,\varsigma}$.

The resulting phase is always a multiple of $\pi/2$, and is remarkably robust against all kinds of disorder. We expand on this in Appendix \ref{sec:Disorder_Ap}.

%#######################################################################################
%#######################################################################################

In this work, we choose the energy imbalance as a control parameter,
but analogous protocols can be implemented introducing a vertical
hopping amplitude between the legs instead, with a Hamiltonian term
of the form $\mathcal{H}_{m}=-\sum_{j=1}^{L}\left(mc_{j,A}^{\dagger}c_{j,B}+h.c.\right)$.
In most experimental implementations, this alternative is more difficult
to implement, and its protection against disorder is equivalent to
the imbalanced case, but we include a discussion on this version of
the models in Appendix \ref{sec:m_version} for completeness, including
their different symmetry classes.

The ICL, especially if implemented with a synthetic dimension (see Section \ref{sec:Exp}), could be used as a basic element in a more complex 2D or 3D structure, in the same way as the SSH chain can be used to build higher order topological insulators \cite{Schindler2018,Liu2019,Dutt2020}. This structure could be used to achieve full connectivity between an arbitrary number of closely packed nodes.

This high connectivity makes the ICL and possible derived models a suitable asset for the implementation of quantum information tasks such as remote quantum gates between external qubits, braiding or entanglement generation in photonic lattices or superconducting circuits, which will be the subject of future work.

\subsection{Fast long-range transfer}

The LR transfer times for single- and two-domain ladders can be obtained analytically with the effective Hamiltonian:

\begin{equation}
t_{\textrm{tr}}^{(N=1)}\!=\!\frac{\pi\epsilon}{2(4J^{2}-\epsilon^{2})}\!\!\left(\frac{2J}{\epsilon}\right)^{\ell+3}\!\!\!=\frac{\pi\epsilon}{2(4J^{2}-\epsilon^{2})}\!\!\left(\frac{2J}{\epsilon}\right)^{\!\!L+1}\label{eq:t_tr_1}
\end{equation}
\begin{equation}
t_{\textrm{tr}}^{(N=2)}\!=\!\frac{\pi\epsilon}{4J^{2}-\epsilon^{2}}\!\!\left(\frac{2J}{\epsilon}\right)^{\ell+2}\!\!\!=\frac{\pi\epsilon}{4J^{2}-\epsilon^{2}}\!\!\left(\frac{2J}{\epsilon}\right)^{\!(L+1)/2}.\label{eq:t_tr_2}
\end{equation}

\begin{figure*}[!tph]
\mbox{%
\includegraphics[width=1\textwidth]{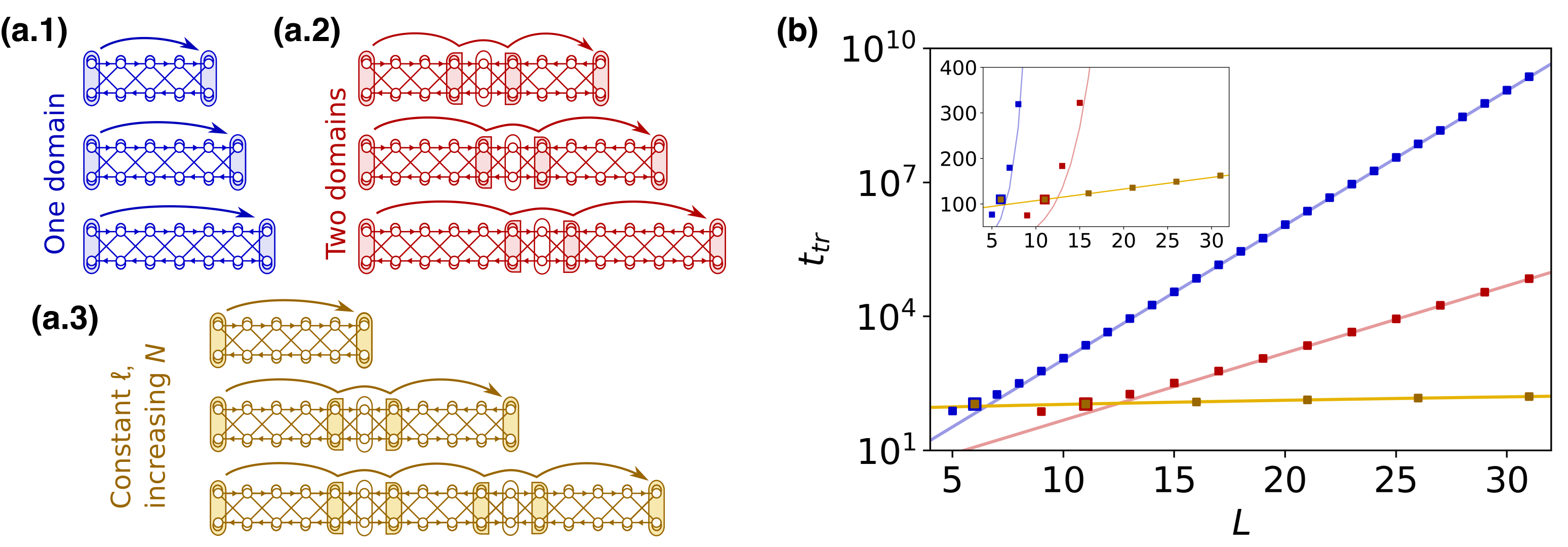}%
}\caption{Transfer time between left and right states in an imbalanced Creutz ladder as
a function of distance. The three cases considered are (a.1) a
single domain of increasing length, in blue, (a.2) two domains of
increasing length, with a single amplifier between them, in red, and
(a.3) an increasing number of domains of length $\ell=4$, in yellow.
(b) Transfer time $t_{\textrm{tr}}$ in the three cases as a function
of the total length of the ladder $L$, color coded as in (a). A logarithmic
scale is used for the vertical axis. The first and second yellow points
coincide with a blue point and a red point, respectively. The simulations
have a maximum control parameter of $\epsilon_{\mathrm{tr}}=1$, a preparation time of $t_{\textrm{prep}}=30$,
and a time step of $\Delta t=0.05$. Analytical results for the transfer
times {[}Eqs. (\ref{eq:t_tr_1},\ref{eq:t_tr_2}){]} are included
in continuous red and blue lines. A linear fit is plotted for the
yellow points. In the inset, the same data is plotted using a linear
vertical scale, in order to appreciate the different trends of the
data. \label{fig:T_tr}}
\end{figure*}

The exponential speed-up found in the SSH chain can also be found in the ICL. We show this in Fig. \ref{fig:T_tr} for single- and two-domain ladders of increasing length, and for ladders of fixed $\ell=4$ and increasing $N$. The details for all protocols are included in Table \ref{tab:Prots}, inside Appendix \ref{sec:kappas}.

Given that the effective model is analogous to the SSH case, the fastest known possible protocols here would also scale linearly with length \cite{Banchi2015,Albanese2004,Christandl2004}.

\subsection{Robustness against disorder}

We also use exact diagonalization to study the behaviour of ICL protocols against symmetry-preserving and general disorder. In the former case, we consider fluctuations
which only depend on the longitudinal coordinate of the ladder $j$,
in order to preserve the chiral symmetry of the ladder $\mathcal{X}_{C}$
(see Appendix \ref{sec:m_version}):

\begin{equation}
\begin{cases}
-J_{j,\sigma}^{(h)} & =-|J+\delta JR_{j}|e^{i\phi/2}\\
-J_{j,\sigma}^{(d)} & =-|J+\delta JR_{j}|,
\end{cases}\label{eq:SPdis}
\end{equation}

where $\delta J$ is the level of off-diagonal disorder, $-J_{j,\sigma}^{(h)}$
and $-J_{j,\sigma}^{(d)}$ are the horizontal and diagonal hopping
terms connecting site $j,\sigma$ to the sites on its right, and $R_{j}\in [-0.5,0.5]$ are random numbers.

For the case with general disorder, we consider:

\begin{equation}
\mu_{j,\sigma}=s_{\sigma}\epsilon+\delta\mu R_{j,\sigma}^{(\mu)}\label{eq:SBdis1}
\end{equation}
\begin{equation}
\begin{cases}
-J_{j,\sigma}^{(h)} & =-|J+\delta JR_{j,\sigma}^{(h)}|e^{i\phi/2}\\
-J_{j,\sigma}^{(d)} & =-|J+\delta JR_{j,\sigma}^{(d)}|
\end{cases},\label{eq:SBdis2}
\end{equation}

where $\delta\mu$ and $\delta J$ represent the level of diagonal
and off-diagonal disorder respectively, $\mu_{j,\sigma}$ is the total
on-site potential on site $j,\sigma$, and all different $R$ variables
are independent random numbers for all different sites and bonds in the interval $[-0.5,0.5]$.

As we explain in Appendix \ref{sec:m_version}, fluctuations in the
control parameters $\epsilon_{j}$ do not break the chiral symmetry
if they are equal for both sites in each rung.

The existence of more degrees of freedom in the Creutz ladder than in the SSH chain for a given $L$ (twice the sites, and four times the number of bonds) have a negative effect on the performance of trivial protocols. To illustrate this and provide a baseline to compare the topological ICL with, we have also included a protocol using potential wells in the trivial Creutz ladder with no magnetic field. Its fidelity with no disorder is  $f=0.95$. We use the same technique as in the trivial chain protocol, which we also include. We use $\mu_0=10$ for both, and tunnellings of $w=J=1$. 

\begin{figure}[!tph]
\mbox{%
\includegraphics[width=1\columnwidth]{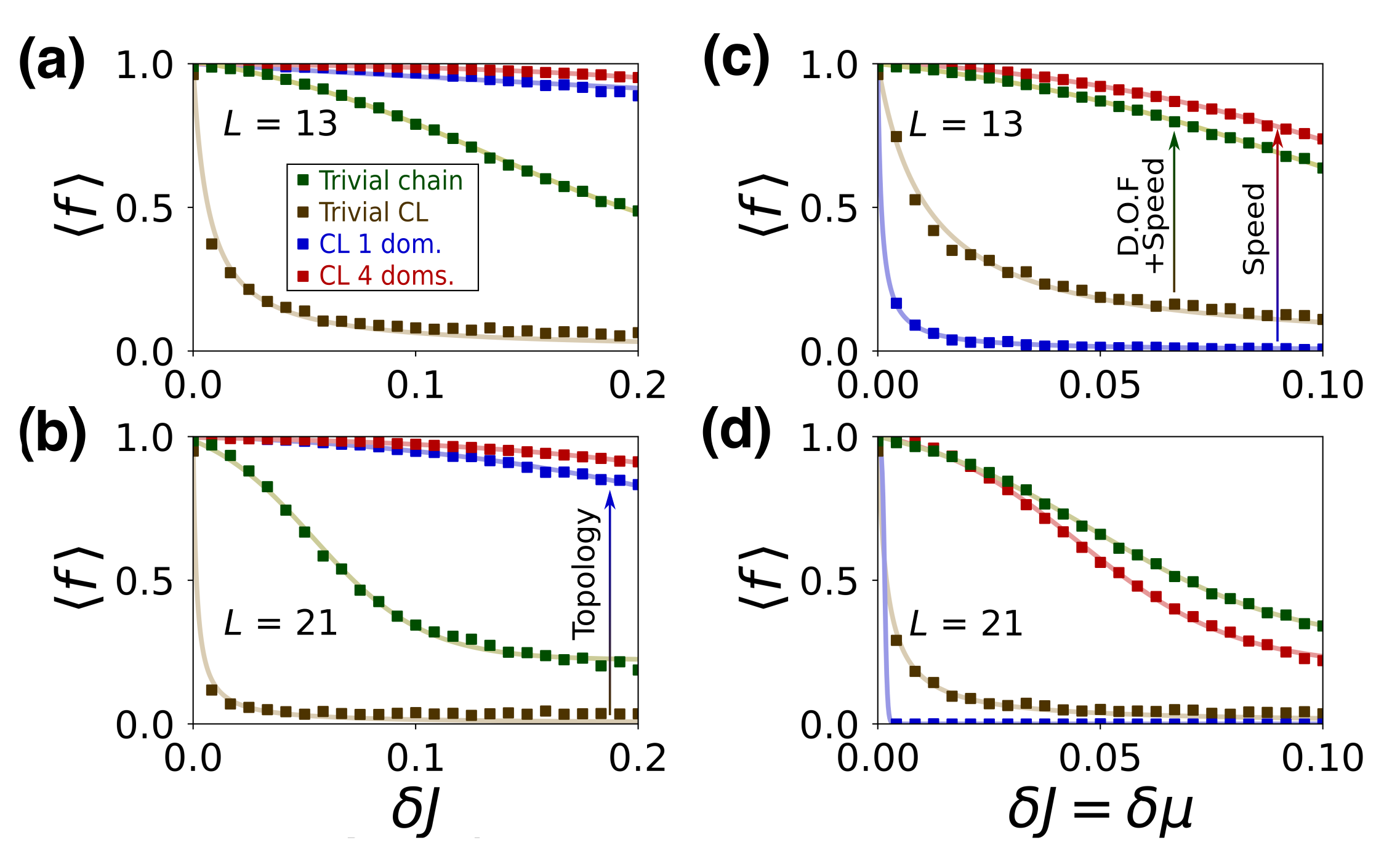}%
}\caption{Fidelity of the transfer protocols against quenched disorder. We compare
trivial transfer protocols in a 1D chain and in a Creutz ladder to
the single- and four-domain topological CL protocols, for a fixed
length $L$. We study the cases $L=13,21$. We consider both off-diagonal
disorder only depending on $j$ (a,b) and general uncorrelated
disorder, both diagonal and off-diagonal (c,d), see main text. The
effect of topological protection can be seen in (a,b), where the topological
protocols show a plateau for low values of disorder. Note the different
horizontal scale for (a,b) and (c,d). In the case with symmetry-breaking
disorder, the multidomain protocol performs much better than the single-domain
one, due to its shorter transfer time. It is comparable with the fidelity
of the trivial chain protocol, which has a much smaller amount of
degrees of freedom that can be affected by disorder. Sigmoid functions
fitted to the data are provided as a guide for the eye. Transfer times
for $L=13$ are $t_{\textrm{tr}}=332.3,3567.7,9013.8,63.0$ for the
trivial chain, the trivial CL and the single- and four-domain CL,
respectively. In the same order, the times for $L=21$ are $t_{\textrm{tr}}=800.5,46739.8,2288679.9,135.5$.\label{fig:Dis}}
\end{figure}

We compare them to LR protocols in single- and four-domain ICLs, using the average over 1000 realizations for all protocols, in systems of lengths $L=13,21$. We show the results in Fig. \ref{fig:Dis}. As can be seen in subfigures (a,b), the topological protocols show a plateau if the chiral symmetry is preserved. For general disorder, (c,d), the four-domain case performs considerably better than the
single-domain one, due to its speed. Its fidelity is then similar to the trivial chain case, even though the latter has a much smaller number of degrees of freedom that can fluctuate. We elaborate on the different factors that affect the fidelity in Appendix \ref{sec:Disorder_Ap}.

%#######################################################################################
%#######################################################################################

%#######################################################################################
%#######################################################################################

\subsection{Experimental proposal \label{sec:Exp}}

As mentioned in the introduction, the imbalanced Creutz ladder can
be implemented using state-of-the-art technology in ultracold atoms
\cite{Zhang2014,Junemann2017,Song2018,HyounKang2020,He2021}, superconducting
circuits \cite{Alaeian2019a,Hung2021} and photonic lattices \cite{Mukherjee2018},
in both waveguide and cavity arrays. For our protocols, it is crucial
that the energy imbalances can be tuned individually, something that
can be easily done in most implementations. Additionally, the Peierls
phases from the synthetic magnetic field must be able to change sign
from one cell to the next in order to implement a domain wall with
a width of a single rung. This is attainable in the photonic and circuit
QED implementations, given that the Peierls phases are obtained with
an on-site energy driving, which can be modified to separately choose
the phase of each bond. On the other hand, the ultracold atoms setup
in \cite{He2021} might be able to engineer a multidomain ladder
if a more complex pattern of standing waves is used, where the resulting
Peierls phase depends on position.

However, the tight-binding model implemented with fermionic atoms
in \cite{Song2018}, with spin acting as a synthetic dimension to
create the two rungs, is especially suited to implement multidomain
Creutz ladders. The tight-binding model that the authors present in
said work is related to the usual Creutz ladder model by the gauge
transformation $c_{j,\sigma}\to\exp[i\pi s_{\sigma}(3/2-j)/2]c_{j,\sigma}$,with
$s_{\sigma}=\delta_{\sigma,A}-\delta_{\sigma,B}$, which only affects
the relative phases between wavefunction components, but not the relevant
physical results. This can be easily seen by examining the magnetic
flux for each closed loop in both lattices, and realizing they are
identical.

\begin{figure}[!tph]
\mbox{%
\includegraphics[width=\columnwidth]{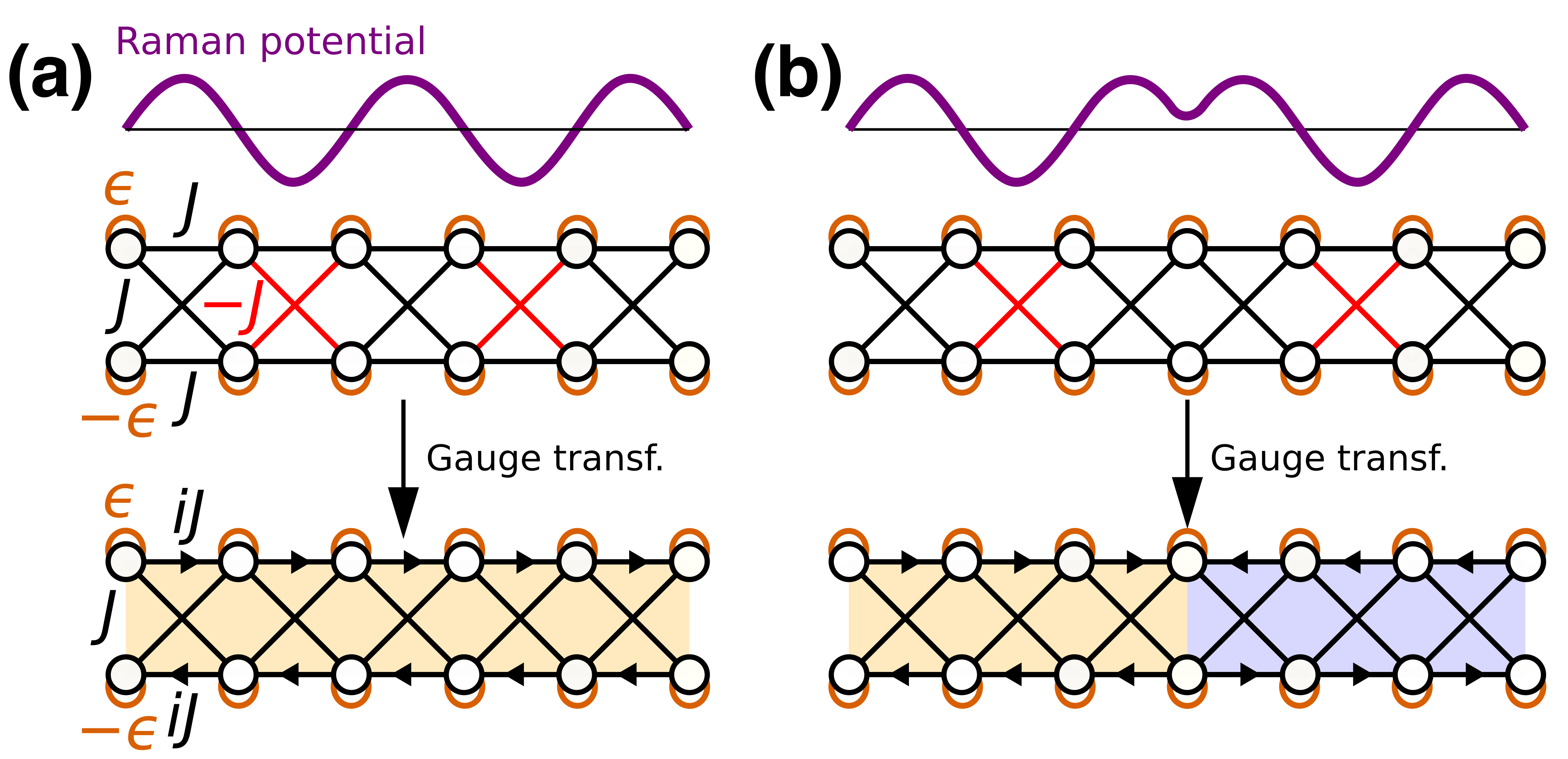}%
}\caption{(a) Effective model for the cold atoms system in \cite{Song2018}, corresponding to a single-domain CL after a gauge transformation. (b) Proposed setup for a two-domain CL. The corresponding Raman potentials are shown in purple. All hopping amplitudes have an additional minus sign.  \label{fig:Exp}}
\end{figure}

The authors build this model using a staggered Raman coupling with
positive and negative terms, which induces a pseudo-spin-orbit term
that creates the diagonal links {[}see Fig. \ref{fig:Exp} (a){]}.
Under this gauge transformation, a system where all even (odd) unit
cells have a negative Raman coupling term corresponds to the $\nu=1\,(-1)$
phase of the Creutz ladder. Thus, the Raman potential can be modified
to create different domains by setting up regions with different alternating
patterns of signs, as shown in Fig. \ref{fig:Exp} (b). This is reminiscent
of the way topological domains are created in the SSH chain. The existence of flat bands in the
Creutz ladder requires the horizontal and diagonal bonds
to be equal in magnitude. A promising aspect of this setup is that
the use of a synthetic dimension will cause noise-induced fluctuations
in the different links of each unit cell to be highly correlated,
which is the kind of disorder which does not break the protecting
chiral symmetry.

The aforementioned gauge transformation can help implement multidomain
Creutz ladders in other platforms in a simpler way, given that it
only uses real hopping amplitudes, as seen in Fig. \ref{fig:Exp}
(b). In photonic lattices, for example, a minus sign can be induced
with an auxilliary waveguide \cite{Kremer2020}, with no need for
a driving protocol.

%#######################################################################################
%#######################################################################################

\subsection{Two-transfer demonstration \label{sec:TwoStateTr}}

\begin{figure}[!tph]
\mbox{%
\includegraphics[width=1\columnwidth]{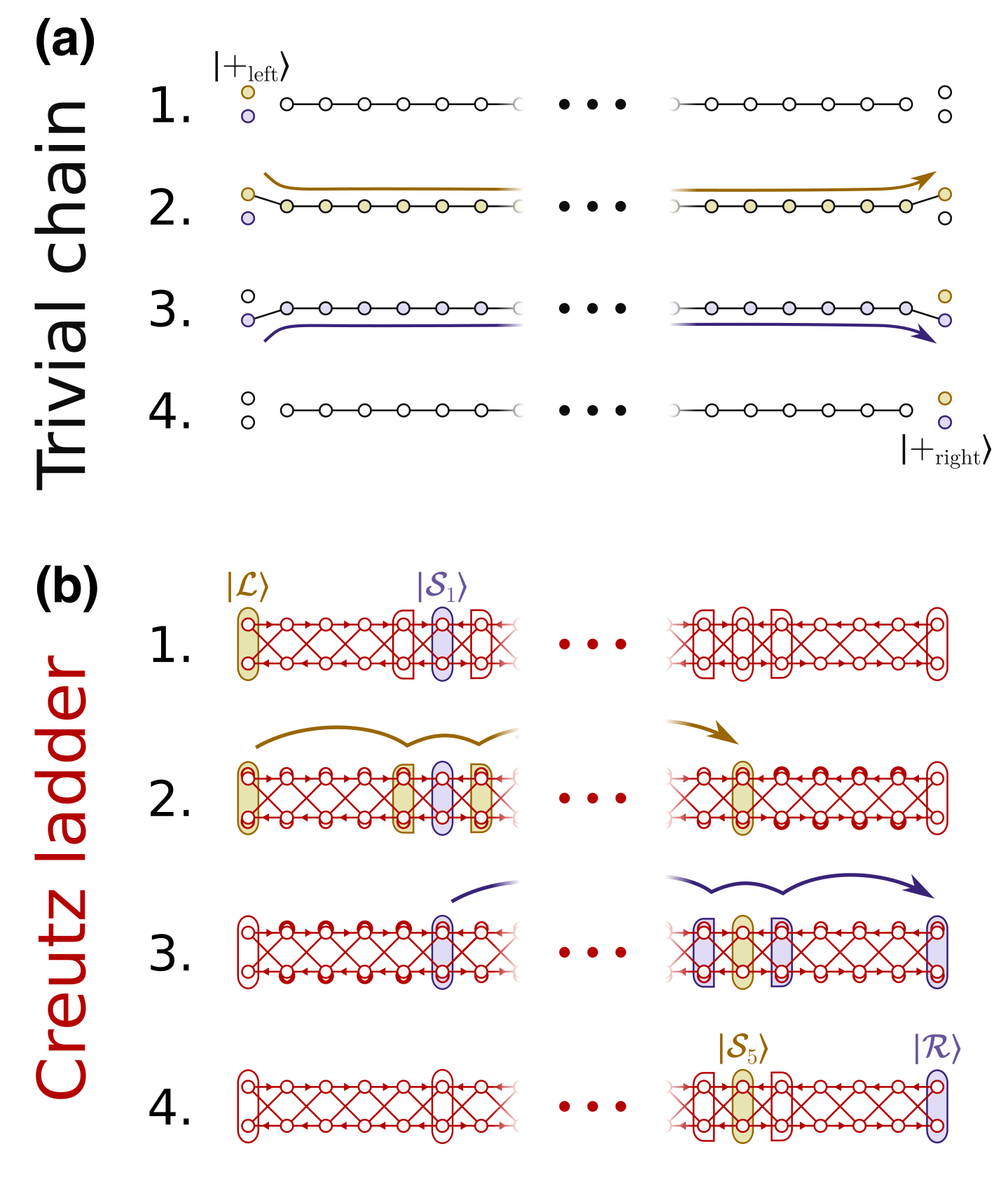}%
}\caption{Transfer of a particle in a superposition of two states. The amplitude
in each of the two states is transferred separately. (a) Transfer
protocol in a trivial chain with $L=26$ for a particle superposed
between two sites. The transfer is induced by using wells of chemical
potential at both edges. (b) Transfer protocol in a six-domain, $L=31$
Creutz ladder for the superposition of two computational states. Each
of the computational states is translated $26$ rungs to the right.
The first transferred state acquires an additional $\pi$ phase in
the CL protocol. \label{fig:QubScheme}}
\end{figure}

Finally, to show an example of a more involved protocol, we consider the transfer of
a superposition of states from the two leftmost to the two rightmost
computational states in a six-domain Creutz ladder with $L=31$:

\begin{align}
\ket{+_{\textrm{left}}^{\textrm{CL}}} & =(\ket{\mathcal{L}}+\ket{\mathcal{S}_{1}})/\sqrt{2}\to\\
\to\ket{\psi_{f}^{\textrm{CL}}} & =\ket{-_{\textrm{right}}^{\textrm{CL}}}=\zeta_{1}\ket{\mathcal{S}_{5}}+\zeta_{2}\ket{\mathcal{R}}.
\end{align}
This is achieved with two successive transfer protocols {[}see Fig.
\ref{fig:QubScheme} (b){]}, and would be impossible in the SSH chain. The acquired phase factors are $\zeta_{1}=-1,\zeta_{2}=1$.
We use $t_{\textrm{prep}}=t_{\textrm{prep}}^{\prime}=30$, $\epsilon_{\textrm{tr}}^{(1)}=J=1$,
$\epsilon_{\textrm{tr}}^{(2)}=\epsilon_{\textrm{tr}}^{(3)}=0.97$,
and a total transfer time of $T_{\textrm{tr}}=391.4$ for the whole
protocol. The fidelity of the transferred state for each system at
the final time $t_{f}$, $F=|\left\langle \psi_{f}^{(\textrm{ideal})}\vert\psi(t_{f})\right\rangle |^{2}$,
is $F=0.996$ for the pristine system.

As a topologically trivial protocol to compare against, we use a transmission
line consisting of a 1D chain of sites with a hopping amplitude of
$J=1$, of the same length as each of the transfers in the CL, and
with two sites at the left ($\ket{1,a/b}$) and right ($\ket{L,a/b}$)
ends {[}see Fig. \ref{fig:QubScheme} (a){]}. The initial state is
$\ket{+_{\textrm{left}}^{\textrm{triv}}}=(\ket{1,a}+\ket{1,b})/\sqrt{2}$,
and two successive transfers are implemented using the same chain,
until the final state, $\ket{\psi_{f}^{\textrm{triv}}}=\ket{+_{\textrm{right}}^{\textrm{triv}}}=(\ket{L,a}+\ket{L,b})/\sqrt{2}$
(plus a certain global phase), is achieved. An initial chemical potential
of $-\mu_{0}=-10J$ is set on the four end sites, and the optimal
transfer time for the full protocol was found to be $T_{tr}=2405.6$. The maximum fidelity obtained in the non-disordered case was $F=0.986$.

\begin{figure}[!tph]
\mbox{%
\includegraphics[width=0.75\columnwidth]{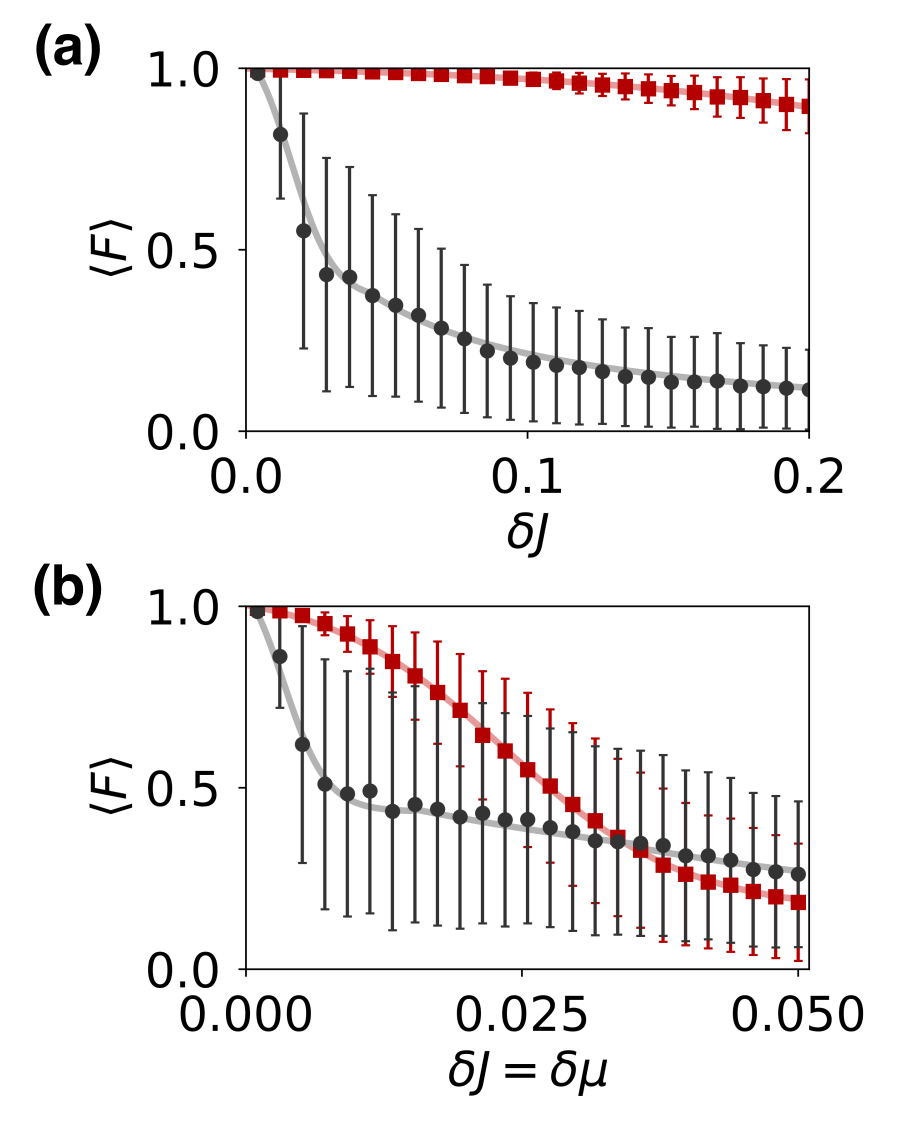}%
}\caption{Average value of the fidelity of the transferred superposed state
over 1000 realizations in the presence of (a) symmetry-preserving
{[}see Eq. (\ref{eq:SPdis}){]} and (b) general disorder {[}Eqs. (\ref{eq:SBdis1},\ref{eq:SBdis2}){]},
using a trivial chain (black) and a six-domain topological CL (red).
Note the different horizontal scale between the figures. We include
its standard deviation as error bars. The CL protocol outperforms
the trivial one by a wide margin in case (a), but it is also better
than it at low levels of disorder in case (b). This is mainly due
to the robustness of the relative phase in the topological protocol
when compared to the trivial one, thanks to its shorter transfer time.
In the trivial case, the relative phase is soon lost, making the fidelity
quickly fall to $\sim0.5$. \label{fig:DisQub}}
\end{figure}

We now obtain the fidelity average over 1000 realizations for both
symmetry-preserving disorder, defined in Eq. (\ref{eq:SPdis}), and
general disorder, see Eqs. (\ref{eq:SBdis1},\ref{eq:SBdis2}). The
results are shown in Fig. \ref{fig:DisQub}. Notice the different disorder scales. The topological protocol
shows a plateau up to disorders of $0.15J$ in the former case, and
still outperforms the trivial protocol in the latter case for low
levels of disorder. The trivial protocol falls quickly to $F=0.5$
in both cases as disorder is turned on, given that the relative phase
of the final superposition---which is dynamical---is almost random even for small disorder values, while the geometrical phases in the topological protocol are more reliable (see Appendix \ref{sec:Disorder_Ap} for more details). In the case of general disorder, the difference is mainly due to the much shorter times of the CL protocol.

As shown, the geometric nature of the relative phase between states in topological insulators can be decisive for protocols involving several transfers. This is also true for setups where external qubits were coupled to the system, as argued in \cite{Lang2017}, given that the acquired phase in the transfer would become relevant then.

\section{Conclusions}

In this work, we propose topologically protected transfer protocols
in the multidomain SSH and Creutz models which exhibit two novel properties.
Firstly, we have obtained an exponential speed-up in our protocols
due to the use of topological domain walls as quantum amplifiers,
when compared to other bidirectional topological transfer protocols.
Protocols of this kind can be useful for the implementation of symmetry-protected
remote quantum gates using topological communication lines, as explored
in \cite{Lang2017}. Our proposal could exponentially accelerate
these operations, even between distant qubits.

Additionally, the magnetic interference and topological properties
of the multidomain Creutz ladder enable all-to-all connectivity between
its boundaries. This could open the realization of long-range symmetry-protected
quantum gates between more than two external qubits, which is an interesting
topic for future work.

Some other possible avenues of research are the detailed analysis
of experimental implementations, including the study of models
with long-range hopping terms and interactions, and the effect
of other types of pulses.

\begin{acknowledgments}
J.Z. would like to thank \'Alvaro G\'omez Le\'on, David Fern\'andez Fern\'andez, Jordi
Pic\'o Cort\'es and Andr\'es Agust\'i Casado for fruitful discussions. The authors want to thank Florent Baboux for identifying an inconsistency in the original manuscript, which allowed for the corrections present in this version. C.E.C. was supported by the
Universidad Complutense de Madrid through Grant No. FEI-EU-19-12.
G.P. and J.Z. were supported by Spain's MINECO through
Grant No. PID2020-117787GB-I00 and by CSIC Research Platform PTI-001.
J.Z. recognizes the FPU program FPU19/03575.
\end{acknowledgments}

\appendix

\section{Adiabatic time evolution and transfer time\label{sec:TimeEvo}}

In this Appendix, we expand on our use of the adiabatic theorem, as
well as perform a step-by-step calculation of the time evolution of the
protected states. All of the considered dynamics are contained within
the invariant subspace around zero energy, $S_{0}$, which is decoupled
from the rest of the Hilbert space due to the topological protection.
In this Appendix, we label all protected states in $S_{0}$ as $\{\ket{i}\}_{i=1}^{M}$.
The evolution of any initial state $\ket{\psi(0)}\in S_{0}$ can then
be expressed as:

\begin{equation}
\ket{\psi(t)}=\sum_{i=1}^{M}\psi_{i}e^{-i\varepsilon_{i}t}\ket{i},
\end{equation}

where $\psi_{i}\equiv\left\langle i\vert\psi(0)\right\rangle$, $M$
is the number of protected states and $\varepsilon_{i}$ are the energies.

This evolution provides the main features of the time evolution, even
when an adiabatic control protocol is added.

As shown by the effective Hamiltonian, which is simply the block of
the Hamiltonian restricted to $S_{0}$, the states involved in the
transfer are isomorphic to a 1D chain, where the exponentially localized
state at each boundary is coupled to those in the two adjacent boundaries.
At the same time, all compact states are decoupled from the rest (for
examples, all $\mathcal{S}$ states in a left-to-right transfer).

For concreteness, let us see some examples in the Creutz ladder (the SSH case is completely equivalent). In a single-domain ladder
with length $L$, the left and right states will hybridize to form
the eigenstates $|\pm\rangle=(|\mathcal{L}\rangle\pm i^{L-1}|\mathcal{R}\rangle)/\sqrt{2}$,
with energies $\pm\varepsilon$. The phase factor $i^{L-1}$ arises
due to the Peierls phases. A straightforward calculation shows that,
if we consider $|\mathcal{L}\rangle$ as the initial state at time
$t=0$, its time evolution will be:

\begin{equation}
|\mathcal{L}(t)\rangle=\cos(\varepsilon t)|\mathcal{L}\rangle-i^{L}\sin(\varepsilon t)|\mathcal{R}\rangle.
\end{equation}

At time $t_{\textrm{tr}}=\pi/(2\varepsilon)$, $|\mathcal{L}(t_{\textrm{tr}})\rangle=-i^{L}|\mathcal{R}\rangle$,
and the particle will have been transferred to the right state, acquiring
a phase that agrees with Eq. (\ref{eq:zetaCL}) (remember that, for
a single domain, $L=\ell+2$). We use this formula for the transfer
time in Eq. (\ref{eq:t_tr_1}), which is obtained by considering a
constant control parameter, for estimating the total transfer time
of the full adiabatic protocol. If we consider $|\mathcal{R}\rangle$
as the initial state instead, we obtain $|\mathcal{R}(t_{\textrm{tr}})\rangle=-i^{-L}|\mathcal{L}\rangle$

For another example, let us consider a two-domain ladder, with an
energy imbalance of $\epsilon$ in all rungs except the central one.
This model has a caged $\mathcal{S}$ state but exponentially localized
$\mathcal{L},\mathcal{R}$ and $\mathcal{P}$ states, which will hybridize
into the states shown in Section \ref{subsec:TopoDomW}, $\ket{\pm}=(\ket{\mathcal{L}}-\ket{\mathcal{R}})/2\pm i^{\ell}\ket{\mathcal{P}}/\sqrt{2}$,
with energies $\pm E$ and $\ket{0_{\mathcal{LR}}}=(\ket{\mathcal{L}}+\ket{\mathcal{R}})/\sqrt{2}$,
at zero energy. The time evolution for the initial state $|\mathcal{L}\rangle$
is:

\begin{equation}
|\mathcal{L}(t)\rangle=\cos^{2}\frac{Et}{2}|\mathcal{L}\rangle+\sin^{2}\frac{Et}{2}|\mathcal{R}\rangle-\frac{i^{\ell+1}}{\sqrt{2}}\sin (Et)|\mathcal{P}\rangle
\end{equation}

At time $t_{\textrm{tr}}=\pi/E$, $|\mathcal{L}(t_{\textrm{tr}})\rangle=|\mathcal{R}\rangle$.
This is the formula we use to obtain Eq. (\ref{eq:t_tr_2}). We can
see that there will be no acquired phase in the gauge we have chosen,
which also agrees with Eq. (\ref{eq:zetaCL}). An initial state of
$|\mathcal{R}\rangle$ would evolve to $|\mathcal{R}(t_{\textrm{tr}})\rangle=|\mathcal{L}\rangle$.

We now consider the adiabatically-changing system, with the pulses
detailed in the main text, and note that there are no level crossings
within $S_{0}$ as long as the control parameter $c$ ($\epsilon,m$
or $v$) is larger than zero. In the $c\to0$ limit, in which we start
and finish the protocol, all states go to $\varepsilon=0$, but they
are all compact with no overlap between them, and so do not mix. Thus,
we can use the adiabatic theorem to predict the time evolution of
the system, as confirmed by our numerical simulations. Each eigenstate
$|i\rangle$ keeps their identity throughout the protocol, although its
energy and wavefunction will change. As our simulations show, no additional
geometric phase is acquired by the localization and delocalization
of the states.

The dynamics described above for the constant-$c$ cases hold for
any $c>0$, and the only difference between different values of $c$
is the energies $\varepsilon_{i}[c(t)]$, and the profile of the states
$|i[c(t)]\rangle$. The time evolution of a given state $\ket{\psi(0)}\in S_{0}$
is now:

\begin{equation}
\ket{\psi(t)}=\sum_{i=1}^{M}\psi_{i}e^{-i\int\varepsilon_{i}[c(t^{\prime})]dt^{\prime}}\ket{i[c(t)]},
\end{equation}

where $\psi_{i}\equiv\left\langle i[c(0)]\vert\psi(0)\right\rangle $.

This explains the similarities between the dynamics in the driven
case and in the constant-$c$ case, only differing slightly in the
transfer time (due to the varying energies) and in the initial and
final wavefunctions, which are compact (and thus, uncoupled to the
rest of the system).

\section{Transfer protocol parameters\label{sec:kappas}}

As detailed in the main text, the protected boundary states at the ends and walls of the models form an effective 1D chain. The left-to-right transfer is reduced to the problem of how to transfer a particle from the first to the last site in this effective 1D chain, the optimal version of which seems to scale linearly with time, as mentioned above. Other intriguing possibilities could be explored here, like  considering the---also especially resilient---adiabatic passage protocol recently proposed in a two-domain SSH chain \cite{Longhi2019a}, or imposing an additional dimerization between the domain walls, in the spirit of \cite{Munoz2018}, to try and obtain an additional degree of protection.

In the latter case, the time dependence is exponential in the effective model length (i.e. the number of domains plus one). These protocols can be faster than those we analyze for some cases, but in principle they will be exponentially slower in the $L\to\infty$ limit.

However, in this work we constrain our study to a simpler version of these protocols, given that it is bidirectional and provides an exponential speed-up over the single-domain case: the Rabi-oscillation-based transfer driven by the modulation of the control parameters over time.

Among all possible ways to modulate them, we have chosen to follow
the pulse form detailed in Eq. (\ref{eq:CParam-1}) for the control
parameter $c$ (which can be $c=v,\epsilon$ or $m$, depending on
the model) simultaneously in all domains, but allowing for a different
value of $c_{\textrm{tr}}$ for each domain. A straightforward numerical
simulation can easily provide the appropriate values of these control
parameters in order to obtain a fidelity of $f>0.995$ in the shortest
time possible, that is, in the first maximum of the last site occupation.
We have considered a value of $c_{\textrm{tr}}^{(1)}$ in the first
and last domain, and then a different value $c_{\textrm{tr}}^{(2)}$
for the second and penultimate, and so on. This, in turn, creates different effective hopping amplitudes in the
effective Hamiltonian.

We include a detailed list of all protocols with $N>2$ that were
used in this work, together with all their parameters, in Table \ref{tab:Prots}.

\begin{table*}[!tph]
\centering
\begin{tabular}{c|c|c|c|c|c|c|c|c|c}
Model & Ref. & $N$ & $\ell$ & $L$ & $c_{\textrm{tr}}^{(1)}$ & $c_{\textrm{tr}}^{(2)}$ & $c_{\textrm{tr}}^{(3)}$ & $t_{\textrm{tr}}$ & $t_{\textrm{prep}}$\tabularnewline
\hline 
SSH & Fig. \ref{fig:SSH_T_tr} & 3 & 4 & 16 & 0.5 & 0.543 & -- & 50.6 & 15\tabularnewline
\hline 
SSH & Fig. \ref{fig:SSH_T_tr} & 4 & 4 & 21 & 0.5 & 0.560 & -- & 55.9 & 15\tabularnewline
\hline 
SSH & Fig. \ref{fig:SSH_T_tr} & 5 & 4 & 26 & 0.5 & 0.561 & 0.566 & 62.2 & 15\tabularnewline
\hline 
SSH & Fig. \ref{fig:SSH_T_tr} & 6 & 4 & 31 & 0.5 & 0.563 & 0.576 & 67.5 & 15\tabularnewline
\hline 
SSH & Fig. \ref{fig:Dis_SSH} & 4 & 2 & 13 & 0.5 & 0.560 & -- & 35.0 & 15\tabularnewline
\hline 
ICL & Fig. \ref{fig:T_tr} & 3 & 4 & 16 & 1 & 0.952 & -- & 123.5 & 30\tabularnewline
\hline 
ICL & Fig. \ref{fig:T_tr} & 4 & 4 & 21 & 1 & 0.969 & -- & 135.7 & 30\tabularnewline
\hline 
ICL & Fig. \ref{fig:T_tr} & 5 & 4 & 26 & 1 & 0.973 & 0.973 & 148.9 & 30\tabularnewline
\hline 
ICL & Fig. \ref{fig:T_tr} & 6 & 4 & 31 & 1 & 0.975 & 0.979 & 162.2 & 30\tabularnewline
\hline 
ICL & Fig. \ref{fig:Dis} & 4 & 2 & 13 & 1 & 0.906 & -- & 63.1 & 30\tabularnewline
\hline 
\end{tabular}\caption{Multidomain transfer protocol parameters. We indicate the model in
which they are implemented (SSH =
SSH chain, ICL = imbalanced Creutz ladder), one of the Figs. which reference them, and the values
of all parameters. The control parameter ($v$ in SSH, $\epsilon$
in ICL) is represented by $c$.\label{tab:Prots}}
\end{table*}

\section{Performance against disorder: key factors and phases \label{sec:Disorder_Ap}}

In this Appendix, we elaborate on the results obtained for the disordered
protocols, by discussing the different factors that can affect the performance
of the protocols.

There are three main factors which determine the robustness against
disorder of a given protocol:
\begin{itemize}
\item The \textbf{topology} of the system. When at least one protecting
symmetry is preserved, the fidelity of the protocol is better in the
topological case, all else being equal. Additionally, the phase of
the wavefunction components is preserved especially well, even if
the protecting symmetries are broken. This is relevant in the CL,
due to the chiralities being defined by the relative phase between
the two legs, and also for more complex protocols, where relative
phases between different computational states carry information.
\item The number of \textbf{degrees of freedom}. More complicated models
like the CL, with a larger amount of moving parts, can induce more
errors in the protocol for the same level of disorder. However, the
relative increase in errors will be platform-dependent, given that
some parameters of the Hamiltonian (and their errors) can be correlated
in some experimental realizations but not in others.
\item The \textbf{speed of the protocol}. Due to the accumulation of errors,
protocols that take more time usually have a lower fidelity than their
faster counterparts, all else being equal.
\end{itemize}
As mentioned in the main text, the value of the acquired phase in
the transfer, $\zeta$, is exceptionally robust in the presence of
disorder. This is because it is a geometric phase, not a dynamical
phase, and so small changes in the transfer time do not affect its
value. We investigate this by representing the circular standard deviation\footnote{The circular standard deviation of a set of data $\{\zeta_{j}\}_{j=1}^{M}$
can be found as $\sigma(\zeta)=\sqrt{2\left(1-\frac{1}{M}\left|\sum_{j=1}^{M}e^{i\zeta_{j}}\right|\right)}$.
It is needed to take into account the periodicity of the phases, and
it ranges between 0 and $\sqrt{2}$. } of the acquired phase values over 1000 realizations.

\begin{figure}[!tph]
\mbox{%
\includegraphics[width=1\columnwidth]{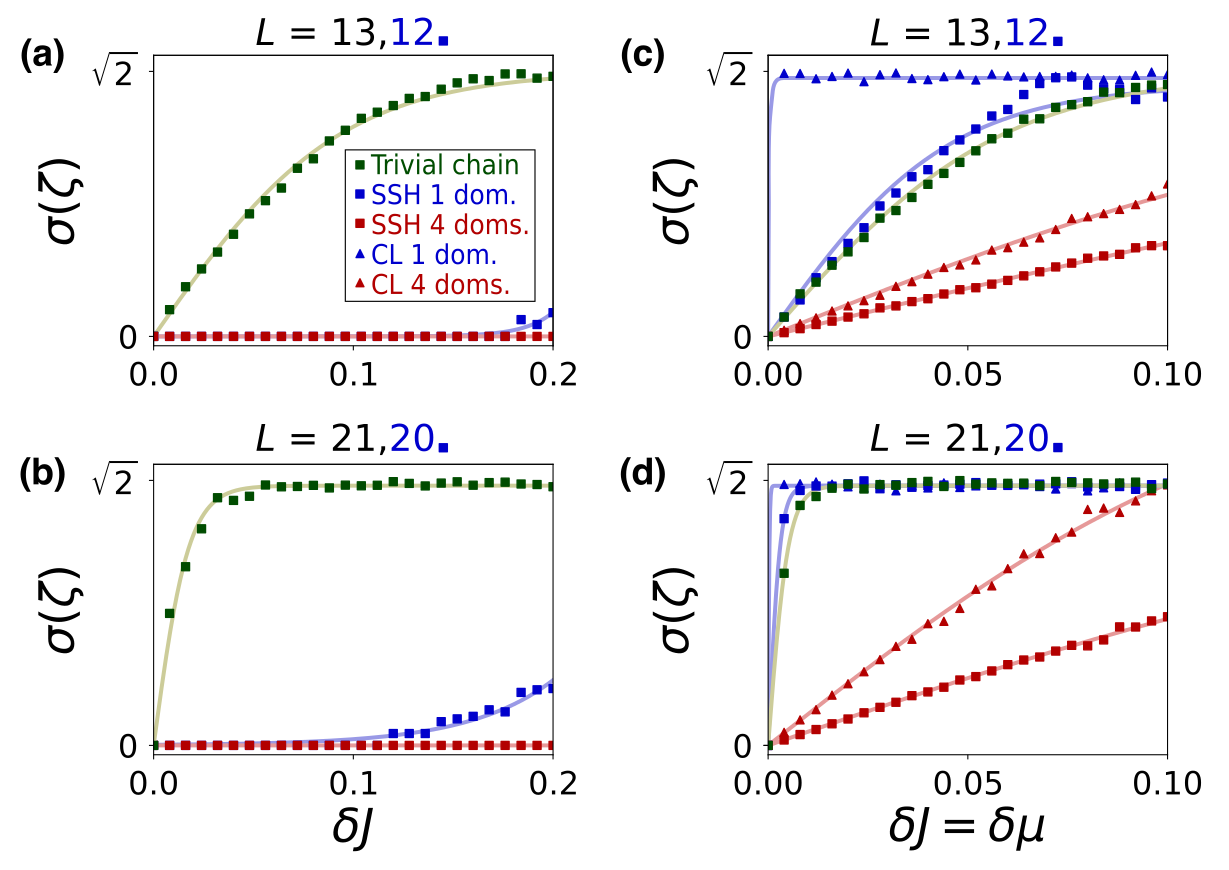}%
}\caption{Standard deviation of the acquired phases over 1000 realizations of
five different protocols, for symmetry-preserving (a,b) and general
(c,d) disorder. The lengths of the models are $L=13\text{ (a,c) and }21$
(b,d) ($L=12,20$ for the single-domain SSH chain). The four topological
protocols show almost perfect results for symmetry-preserving disorder
up to $0.2J$. In the case of general disorder, the four-domain protocols
show a clear advantage over the rest, due
to their shorter transfer times. Continuous lines fitted to the points
are added as a guide for the eye. \label{fig:Dis_Ph}}
\end{figure}

In the case of symmetry-preserving disorder, the acquired phases are
almost completely unperturbed for disorder strengths of more than
20\% of the energy scale of the model in both the SSH chain and CL
protocols, with only the single-domain SSH protocol showing the effects
of disorder near $\delta J\simeq 0.1$. This protection does not seem
to carry over to the general disorder case, but we see a better performance
in the faster, four-domain SSH and Creutz models than in a single-domain
or trivial protocol, even in the case where the fidelity in the CL
did not outperform the trivial chain, compare Fig. \ref{fig:Dis_Ph}
(d) with Fig. \ref{fig:Dis} (d). In general, the stability of the
acquired phase seems to outlast the fidelity plateau as disorder increases.

This behaviour is useful in situations where the predictability of
the phase is desired, like quantum information implementations in
which the topological model is coupled to external qubits (e.g., if
used as a photonic communication line to implement remote quantum
gates, like in \cite{Lang2017}), or more complex transfer protocols
where relative phases between boundary states also have to be preserved.

% \section{The multidomain SSH chain effective Hamiltonian\label{sec:SSH_Eff}}

%The effective Hamiltonian for the boundary states
%in the SSH chain has the same form as Eq. (\ref{eq:Heff}), but substituting
%all $\mathcal{P}$ states \textendash which do not exist here\textendash{}
%by \emph{$\mathcal{S}$} states. The effective hopping amplitudes
%now have the form:
%
%\begin{align}
%t_{1} & =v\mathcal{N_{L}^{\prime}}\mathcal{N}_{\mathcal{S}_{1}}^{\prime}\left(-\frac{w}{v}\right)^{-\ell/2-1}\\
%t_{k} & =-v\mathcal{N}_{\mathcal{S}_{k}}^{\prime}\mathcal{N}_{\mathcal{S}_{k-1}}^{\prime}\left(-\frac{w}{v}\right)^{-\ell/2},\,\,\,\,\,k=2,\ldots,N-1\\
%t_{N} & =v\mathcal{N_{R}^{\prime}}\mathcal{N}_{\mathcal{S}_{N-1}}^{\prime}\left(-\frac{w}{v}\right)^{-\ell/2-1},
%\end{align}
%
%where the normalization constants can be approximated as:

%\begin{align}
%\mathcal{N}_{\mathcal{L}}^{\mathcal{\prime}} & =\mathcal{N_{R}^{\prime}}=\sqrt{w^{2}/v^{2}-1}\\
%\mathcal{N}_{\mathcal{S}_{k}}^{\prime} & =\sqrt{\frac{w^{2}-v^{2}}{w^{2}+v^{2}}}\,\,\,\,\,\,\,\forall k.
%\end{align}

\section{Topological state preparation in the Creutz ladder\label{sec:CSSH}}

\begin{figure}[!tph]
\mbox{%
\includegraphics[width=1\columnwidth]{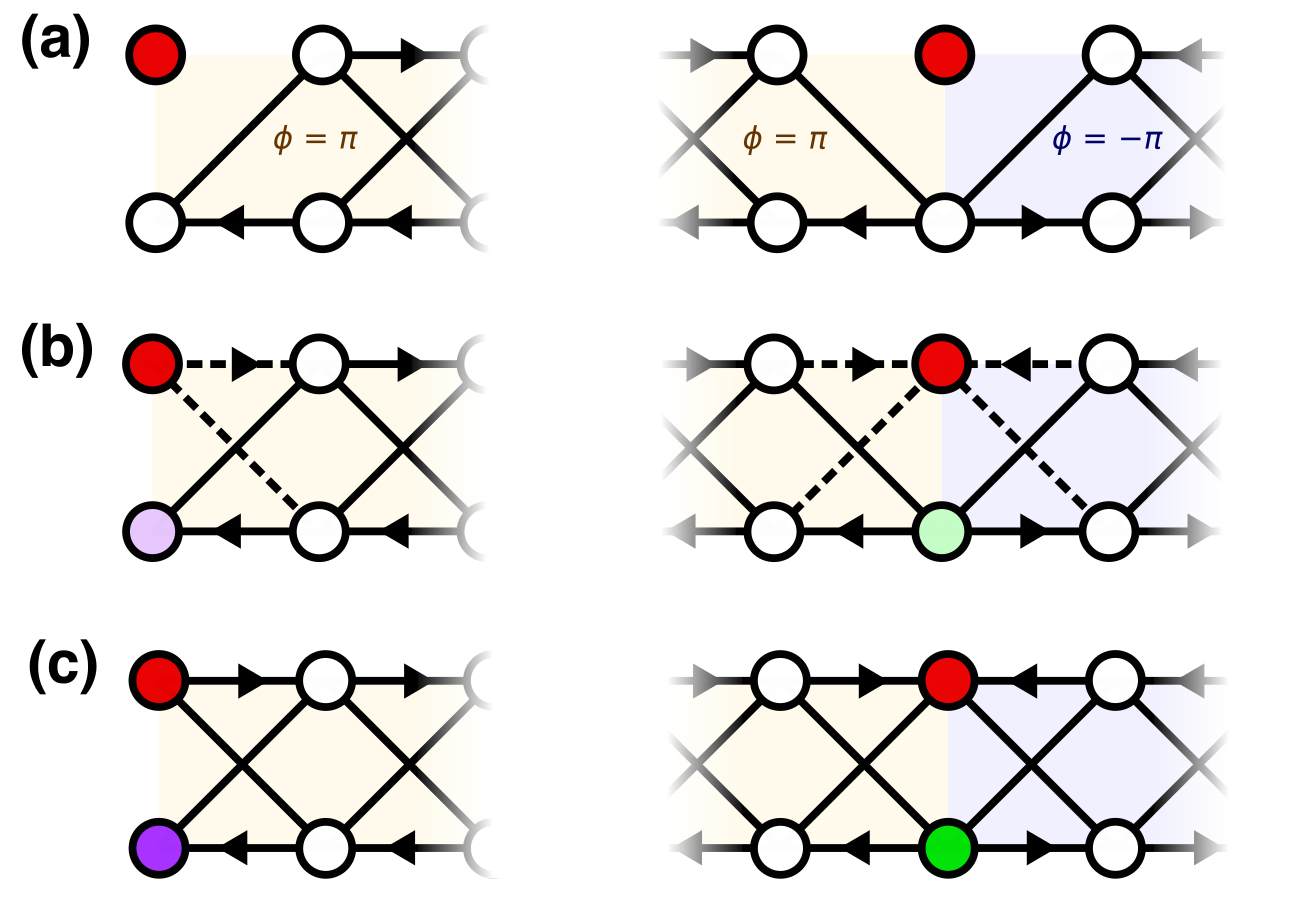}%
}\caption{Preparation of the computational states $\ket{\mathcal{L}}$ (left
of the figure) and $\ket{\mathcal{S}_{1}}$ (right) in the Creutz
ladder. (a) The initial state has a particle localized in a single
site in leg $A$. (b) The hopping amplitudes connected to that site
(dashed lines) must start at a value of zero, and then be adiabatically
switched on, with the proper complex phases. This protocol is topologically
protected by chiral symmetry, given that each intermediate state corresponds
to a topological state in a CSSH ladder \cite{Zurita2021}, to which
the system is locally equivalent. (c) The final state is the corresponding
computational state in the Creutz ladder. Magnetic interference will
cause the site in leg $B$ to acquire the necessary phase, represented
by color. Red is used for phase $0$, green for phase $\pi/2$ and
purple for phase $-\pi/2$. \label{fig:CSSH}}
\end{figure}

The left, right and $\mathcal{S}$-type topological states of the
rungless Creutz ladder, which are localized in two sites, can be prepared
starting from a particle confined to a single site, while retaining
topological protection. To do this, the hopping amplitudes of the
given site with its neighbours must start off at zero, and then be
turned on adiabatically until they reach the value of the rest
of the hopping amplitudes, with the appropriate complex phases (see Fig. \ref{fig:CSSH}). The
symmetry protection is retained during this process, because the state
corresponds to a topological state of a rhomboid CSSH ladder of varying
$J/J^{\prime}$ \cite{Zurita2021}. Although the full system is not
equivalent to the CSSH ladder during the preparation, it is locally
equivalent to it around the computational states. Given the localized
nature of all eigenstates during the protocol due to AB caging, this
local equivalence is enough to make the computational state completely
analogous to its CSSH ladder counterpart, including the topological
protection, so the preparation protocol will be symmetry-protected
against off-diagonal disorder.

\section{Derivation of the protected zero modes\label{sec:DerivStates}}

In order to obtain the form of the protected states, we consider a
domain wall between two semi-infinite domains, with $\nu=\pm1$ for
the left/right domain. The energy imbalance at the domain wall is
set to zero, while it is equal to $\epsilon$ for all other rungs.
We also set the origin of coordinates at the domain wall. Then, it
can be easily checked that state $\ket{\mathcal{S}}=[\ket{0,A}+i\ket{0,B}]/\sqrt{2}$
is an eigenstate with zero energy.

Let us now consider the following ansatz:

\begin{equation}
\ket{\mathcal{P}}=\mathcal{N}\sum_{\forall j\ne0}\zeta_{j}e^{-\alpha\left|j\right|}\ket{j,+},\label{eq:ansatz}
\end{equation}

where $\mathcal{N}$ is a normalization constant, $\zeta_{j}$ are
phase factors of magnitude one, $\alpha$ is an unknown coefficient,
and $\ket{\pm}=\ket{A}\pm i\ket{B}$ are the chiral states.

We consider an ansatz with positive chirality due to the sign of the
difference between the invariants of the two domains. We did not include
any component with $j=0$, given that state $|\mathcal{S}\rangle=|0,+\rangle$
is itself a different eigenstate. We want to check if ansatz (\ref{eq:ansatz})
can satisfy the eigenstate equation, $\mathcal{H}\ket{\mathcal{P}}=\varepsilon\ket{\mathcal{P}}$,
for some values of $\alpha,\zeta_{j}$ and $\varepsilon$.

Due to the chiral symmetry of the Hamiltonian, $\{\mathcal{X},\mathcal{H}\}=0$,
its application on a chiral state like $\ket{\mathcal{P}}$ will give
as a result a state with the opposite chirality. This can be seen
using partial inner products in the following way:

\begin{equation}
\bra{+}\mathcal{H}\ket{+}=-\bra{+}\mathcal{X^{\dagger}HX}\ket{+}=-\bra{+}\mathcal{H}\ket{+}=0.
\end{equation}

Given that $\ket{\pm}$ are a complete basis of the internal space,
$\mathcal{H}\ket{+}$ must be chiral too, with a negative chirality.

For this reason, the eigenstate equation can only be satisfied for
$\varepsilon=0$, so all coefficients of state $\mathcal{H}\ket{\mathcal{P}}$
must vanish. A simple calculation yields:

\begin{equation*}
\mathcal{H}\ket{\mathcal{P}}=\mathcal{N}\sum_{j\ne0}\left[-2Ji\zeta_{j^{\prime}}e^{-\alpha|j+1|}+\zeta_{j}e^{-\alpha|j|}\epsilon\right]|j,-\rangle+
\end{equation*}
\begin{equation}
-\mathcal{N}Ji(\zeta_{1}+\zeta_{-1})e^{-\alpha}|0,-\rangle
\end{equation}

where $j^{\prime}=\mathrm{(sign}j)(|j|+1)$, which implies:

\begin{equation}
\zeta_{1}+\zeta_{-1}=0
\end{equation}

\begin{equation}
e^{-\alpha}=-i\frac{\zeta_{j}}{\zeta_{j^{\prime}}}\frac{\epsilon}{2J},\quad j\ne0.
\end{equation}

Given that $\alpha\in\mathbb{R}^{+}$ by construction, we get that
the phase factors follow the rules: 

\begin{equation}
\begin{cases}
\zeta_{j} & =-i\zeta_{j-1}\quad j<0\\
\zeta_{j+1} & =i\zeta_{j}\quad j>0\\
\zeta_{1} & =-\zeta_{-1}
\end{cases}
\end{equation}

We choose the gauge in which $\zeta_{-1}=1$, and so $\zeta_{j}=[-i\,\mathrm{sign}(j)]^{j+1}$,
which takes the cyclic values $\pm1,\pm i$.

On the other hand, the exponential coefficient is $\alpha=\log(2J/\epsilon)$
and the $\mathcal{P}$ state can be written for the first wall, after
the appropriate change of coordinates, as in Eq. (\ref{eq:Pm-1-1}).
Completely analogous calculations yield the formulas for the rest
of the protected states of the multidomain Creutz and SSH models.

\section{Runged vs. imbalanced Creutz ladder: symmetries and states\label{sec:m_version}}

In this Appendix, we discuss the differences and similarities between
the imbalanced ($\epsilon\ne0,m=0$) and the runged ($m\ne0,\epsilon=0$)
Creutz ladders.

\begin{figure}[!tph]
\mbox{%
\includegraphics[width=0.8\columnwidth]{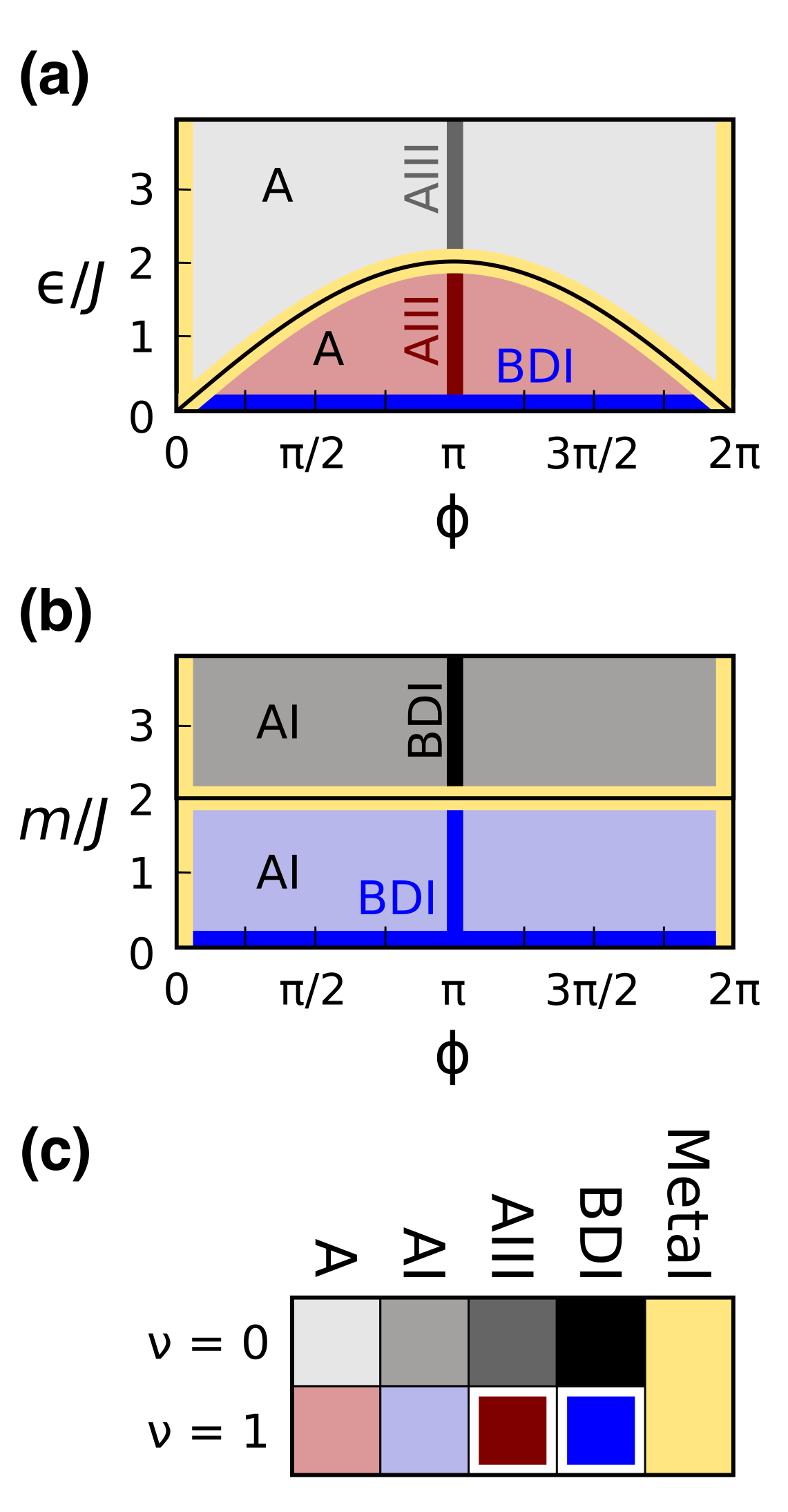}%
}\caption{Phase diagrams showing the symmetry classes and winding numbers of
the two Creutz ladder variants. (a) Imbalanced Creutz ladder phase
diagram, with $m=0$. (b) Runged Creutz ladder phase diagram, with
$\epsilon=0$. (c) Color code for the symmetry classes and winding
number values ($\text{\ensuremath{\nu}}$). Topological phases explained
by the Altland-Zirnbauer classification \cite{Altland1997} are indicated
by a white outline. \label{fig:SymCl}}
\end{figure}

The rungless, balanced regime ($m=\epsilon=0$) of the CL belongs
to the BDI class as long as $\phi\ne0\mod2\pi$, with a hidden chiral
symmetry that can be expressed as $\mathcal{X}_{S}^{(4)}=\textrm{diag}(\uno_{2},-\mathbb{\uno}_{2})$
if we choose a four-site unit cell \cite{Zurita2021}. The runged
case also belongs to the BDI class if $\phi=\pi\mod2\pi$, with different
symmetries than in the previous case. The chiral symmetry is then
$\mathcal{X}_{C}=i\sigma_{y}$. The imbalanced CL belongs to the AIII
class if $\phi=\pi\mod2\pi$, with the same chiral symmetry $\mathcal{X}_{C}$.
When either of the control parameters ($m$ or $\epsilon$) is nonzero and $\phi\ne0,\pi\mod2\pi$, the system belongs to one of the nontopological symmetry classes in 1D. Topological edge states can still present in the model, but they are related only to crystalline symmetries instead \cite{Li2015}. The phase diagrams for the imbalanced and runged Creutz ladders are shown in Fig. \ref{fig:SymCl}. The main difference between them is that the topological region is $m<2J$ for the balanced case, but $\epsilon<2J\sin(\phi/2)$ for the rungless case. However, the two chiral regions, where the states are protected, are the same.

Topological states are protected by these two chiral symmetries, but not
for all types of disorder. Diagonal (on-site) disorder always breaks
the symmetries of the models, while each of the symmetries protects
against different kinds of off-diagonal disorder, as shown in Table
\ref{tab:syms}. The system is only topologically protected if the
off-diagonal disorder does not depend on the internal coordinate $\sigma$.
For this reason, implementations that use a synthetic dimension for
the two legs will most likely be advantageous over those using two
real dimensions, given that the inter-cell parameters will be more
correlated in the former case.

Crucially, fluctuations in any of the control parameters ($m$ or
$\epsilon$) preserve the chiral symmetry $\mathcal{X}_{C}$, and
thus the topological protection.

\begin{table}[!tph]
\begin{tabular}{>{\footnotesize\centering}p{1.2cm}|c|c|c|c|c}
 & \!$\delta J_{h/d}(j,\sigma)$\! \!\!& \!$\delta  J \!(j)$ & \!$\delta\phi(j,\sigma)$ & \!$\delta c(j)$ & \!$\!\delta\mu^{*}$\tabularnewline
\hline 
\enspace $\mathcal{X}_{C}$ {[}$\text{\ensuremath{\phi=\pm\pi}}${]} &  & $\checkmark$ &  & $\checkmark$ & \tabularnewline
\hline 
\enspace $\mathcal{X}_{S}^{(4)}$
{[}$\text{\ensuremath{m\!=\!\epsilon\!=\!0}}${]} & $\checkmark$ & $\checkmark$ & $\checkmark$ &  & \tabularnewline
\end{tabular}

\caption{Types of disorder and the chiral symmetries in the Creutz ladder that
are preserved ($\checkmark$) or broken by them. They are defined as $\mathcal{X}_{C}=\sigma_{y}$, $\mathcal{X}_{S}^{(4)}=\textrm{diag}(\uno_{2},-\mathbb{\uno}_{2})$. The regime in which
the symmetry is present is shown in square brackets. The dependence
of the disorder terms is indicated in parentheses. $\delta J_{h/d}(j,\sigma)$
allows for fluctuations of different value in each of the four inter-cell
links, while $\delta J(j)$ means that we apply the same fluctuation
to all four inter-cell hopping terms. $\delta c(j)$ represents a
rung-dependent disorder in one of the control parameters: the vertical
links ($c=m$) or the energy imbalance $(c=\epsilon$), $\delta\phi(j,\sigma)$
indicates a bond-dependent fluctuation of the Peierls phases, and
$\delta\mu^{*}$ stands for a general non-homogeneous fluctuation
in the chemical potentials. For a detailed account of all symmetries,
cf. \cite{Zurita2021}.\label{tab:syms}}
\end{table}

The runged multidomain CL with rungless domain walls has the following
topological states:
\begin{widetext}
\begin{align}
\ket{\mathcal{L}} & =-\mathcal{N_{L}}\sum_{j=1}^{\ell+1}\left(-\frac{2J}{m}\right)^{-j}\ket{j}\otimes\left(\begin{array}{c}
1\\
-i
\end{array}\right)\label{eq:Lm-2}\\
\ket{\mathcal{R}} & =-\mathcal{N_{R}}\sum_{j=L-\ell}^{L}\left(-\frac{2J}{m}\right)^{j-L-1}\ket{j}\otimes\left(\begin{array}{c}
1\\
(-1)^{N+1}i
\end{array}\right)\label{Rm-2}\\
\ket{\mathcal{P}_{k}} & =-\mathcal{N}_{\mathcal{P}_{k}}\sum_{j=(k-1)(\ell+1)+2}^{k(\ell+1)}\left(-\frac{2J}{m}\right)^{j-k(\ell+1)-1}\ket{j}\otimes\left(\begin{array}{c}
1\\
(-1)^{k+1}i
\end{array}\right)\nonumber \\
 & +\mathcal{N}_{\mathcal{P}_{k}}\sum_{j=k(\ell+1)+2}^{(k+1)(\ell+1)}\left(-\frac{2J}{m}\right)\ket{j}\otimes\left(\begin{array}{c}
1\\
(-1)^{k+1}i
\end{array}\right),\label{eq:Pm-2}
\end{align}

and its effective Hamiltonian has the same form as the imbalanced
case, with the following hopping amplitudes:

\begin{equation}
v_{k}=2(-1)^{k+p_{k-1}}im\mathcal{N}_{k}\mathcal{N}_{k-1}\left(-\frac{2J}{m}\right){}^{-d-2},
\end{equation}

where $d$ is the distance between the maxima of the involved states,
and $p_{k-1}=0$ if the $(k-1)$-th state is a \emph{$\mathcal{P}$
}state, and $p_{k-1}=1$ otherwise.

In a transfer involving an $\mathcal{S}$ state that extends to the
left or right, this state will take the form:

\begin{equation}
\ket{\mathcal{S}_{k}}_{\mathrm{left}}=\mathcal{N}_{\mathcal{S}_{k}}\sum_{j=(k-1)(\ell+1)+2}^{k(\ell+1)+1}\left(-\frac{2J}{m}\right)^{j-k(\ell+1)-1}\ket{j}\otimes\left(\begin{array}{c}
1\\
(-1)^{k+1}i
\end{array}\right)
\end{equation}
\begin{equation}
\ket{\mathcal{S}_{k}}_{\mathrm{right}}=\mathcal{N}_{\mathcal{S}_{k}}\sum_{j=k(\ell+1)+1}^{(k+1)(\ell+1)}\left(-\frac{2J}{m}\right)^{k(\ell+1)-j+1}\ket{j}\otimes\left(\begin{array}{c}
1\\
(-1)^{k+1}i
\end{array}\right).
\end{equation}
\end{widetext}

The acquired phases in a transfer jumping over $n_{w}$ walls with
a domain length of $\ell$ are different from the imbalanced case,
and can only be $0$ or $\pi$, making it easier to compensate if
needed for quantum information applications:

\begin{equation}
\zeta(\ell,n_{w},x,\varsigma)=\begin{cases}
(-1)^{\ell+n_{w}/2+\delta_{-x,\varsigma}} & \textrm{for even }n_{w}\\
(-1)^{(n_{w}-1)/2} & \textrm{for odd }n_{w},
\end{cases}
\end{equation}

where $x=\pm1$ is the chirality of the leftmost transferred state,
$\varsigma$ is the direction of transfer ($\pm1$ for left-to-right/right-to-left).

Apart from the phases, the rest of the dynamics of the model are identical to those in the imbalanced CL, if we remain in the AZ-topological phases. In particular, all transfer times are the same, described by Eqs. (\ref{eq:t_tr_1},\ref{eq:t_tr_2}), with $m$ playing the part of $\epsilon$. This was confirmed both analytically and numerically.

%\bibliographystyle{quantum}
%\bibliography{BibliographyFixed}

\end{document}